\documentclass{aa}  

\usepackage{txfonts}
\usepackage[colorlinks=true,citecolor=blue,urlcolor=blue,breaklinks=true]{hyperref}
\usepackage{graphicx,epstopdf}
\DeclareGraphicsExtensions{.pdf}
\begin{document} 

\title{Jets in FR0 radio galaxies}


   \author{G.~Giovannini \inst{1,2} \and
          R.~D.~Baldi \inst{1} \and A.~Capetti \inst{3} \and
          M.~Giroletti \inst{1} \and R.~Lico \inst{4,1}
          }
          
\institute{INAF - Istituto di Radioastronomia, via Gobetti 101 40129 Bologna\\
              \email{ggiovann@ira.inaf.it}
         \and
Dipartimento di Fisica e Astronomia - University of Bologna, via Gobetti 93/2 
         \and
         INAF - Osservatorio Astronomico di Torino
         \and
         Instituto de Astrof\'{\i}sica de Andaluc\'{\i}a-CSIC, Glorieta de la Astronom\'{\i}a s/n, 18008 Granada, Spain. }
         
   \date{Received .......; accepted ......}

 
  \abstract
   {The local radio-loud AGN population is dominated by compact sources named FR0s. These sources show features, for example the host type, the mass of the supermassive black hole (SMBH), and the multi-band nuclear characteristics, that are similar to those of FRI radio galaxies. However, in the radio band, while FR0 and FRI share the same nuclear properties, the kiloparsec-scale diffuse component dominant in FRI is missing in FR0s. Previous very-long-baseline interferometry (VLBI) observations of a small sample of FR0s show a complex structure, mostly symmetric (two-sided jets) with respect to the central core.}
   {With this project we would like to study the parsec-scale structure in FR0s in comparison with that of FRI sources. Jets in FRI are relativistic on the parsec scale and decrease their velocity becoming subrelativistic on the kiloparsec scale. We would like to test whether this result also applies  to the jets in FR0s or, alternatively, whether they are subrelativistic on the parsec scale. This might be the reason why they are unable to grow, because of instabilities, related to a low jet bulk velocity. }
   {To this end we observed 18 FR0 galaxies with the VLBA at 1.5 and 5 GHz and/or with the EVN at 1.7 GHz  and produced detailed images at milliarcsec resolution of their nuclear emission to study the jet and core structure.}
   {All sources have been detected but one. Four sources are unresolved, even in these high-resolution images; jets have been detected in all other sources. We derived the distribution of the jet-to-counter-jet ratio of FR0s and found that it is significantly different from that of FRIs, suggesting different jet bulk speed velocities. }
   {Combining the present data with published data of FR0 with VLBI observations, we derive that the radio structure of FR0 galaxies shows  strong
 evidence that parsec-scale jets in FR0 sources are mildly  
 relativistic with a bulk velocity on the order of 0.5c or less. A jet structure with a thin inner relativistic spine surrounded by a low-velocity sheath could be in agreement with the SMBH and jet launch region properties.}

               
   \maketitle
%
\section{Introduction}
It is now clear that most of the radio sources associated with
low-redshift galaxies (e.g., \citealt{heckman14}) are compact, with corresponding linear sizes
smaller than $\sim$10 kpc \citep{baldi09,baldi10b}.
These compact sources are located in red massive
($\sim$10$^{11}$ solar masses) early-type
galaxies, with high supermassive black hole (SMBH) masses ($\sim$ 10$^8 M_\odot$), and
spectroscopically classified
as low-excitation radio galaxies (LERGs).
They also show the same correlation between radio core power and [O III] line luminosity of FRIs, but
are more core-dominated (by a factor of $\sim$ 30) than
FRIs and show a clear deficit of extended radio emission.
For these reasons these compact sources were named FR0 \citep{ghisellini11,sadler14,baldi15}, being similar to FRI from the optical
and nuclear radio emission point of view, but without an extended
emission dominant at low frequencies.

The available information from observations of FR0s is
still limited, even in the radio band, but recently major results have
been obtained (see, e.g., JVLA observations presented in \citealt{baldi15,baldi19}).
Most of the sources appear unresolved down to an angular resolution of $\sim$ 0.3 arcsec, which corresponds to 100–300 pc. A few sources show radio emission extended over a few arcseconds, on a scale of 2–14 kpc.

From an X-ray study of a sample of 19 FR0 radio galaxies \citet{torresi18} find
that the X-ray photons are likely produced by the jet, as attested by the
observed correlation between X-ray (2–10 keV) and radio (5 GHz)
luminosities, similar to FRI radio galaxies.  Overall, they find
that the X-ray properties of FR0s are indistinguishable from those of FRIs.
A comparison between FR0s and low-luminosity BL Lacs rules out important
beaming effects in the X-ray emission of the compact radio galaxies.

A complete sample of FR0 radio galaxies have been selected from radio and optical surveys, the  FR0CAT catalog (FR0CAT, \citealt{baldi18}) to constitute a workbench sample for the following studies of these compact radio-loud AGN.
 Thanks to LOFAR  and GMRT observations of FR0 galaxies from the FR0CAT, \citet{capetti19,capetti20}  conclude that the fraction of FR0s showing evidence for the presence of jets at low radio frequencies, by including both spectral and morphological information,
is at least $\sim$40\%. This study confirms that FR0s and
FRIs can be interpreted as two extremes of a continuous population of jetted
sources \citep{capetti17a,baldi18}, with the FR0s representing the low end in size and radio power.

Studying the properties and the origin of the radio
emission in the most luminous, early-type galaxies (ETGs) in the nearby
Universe with LOFAR observations, \citet{capetti22}  found that about two-thirds of
the detected ETGs are unresolved, with sizes on the order of 4 kpc or lower, confirming the prevalence of compact radio sources in local universe. Because these observations are at low radio frequencies, they exclude the presence of extended steep spectrum lobes and the authors  conclude that these sources are FR0 candidates.

Various evolutionary scenarios were investigated
(e.g., \citealt{baldi19,baldi19rev}).
First, the possibility that
all FR0s are young sources eventually evolving into extended sources was
ruled out by the distribution of radio sizes.
Similarly, a time-dependent scenario where  accretion
or jet launching properties prevent the formation of large-scale radio
structures appears to be implausible, owing to the large abundance of subkiloparsec
objects.
Finally, a scenario in which FR0s are produced by mildly relativistic jets is
consistent with the data, but requires observations of a larger sample
to be properly tested.

\citet{garofalo10} discussed a model where powerful jets are
due to fast spin retrograde accreting SMBHs, while a prograde configuration is identified with radio quiet galaxies. The natural time evolution of fast spin retrograde SMBHs is toward 
low-spin retrograde SMBHs followed by a prograde configuration \citep{garofalo16}. 
\citet{garofalo19} identified a region of the jet power versus spin space corresponding to relatively weak jets (see their Fig. 1), possibly identified with FR~0 jets properties. Prograde low-spin BHs can shape compact and/or weak FR~0 jets.

\citet{massaro20} compared the properties of the large-scale
environment, an orientation-independent property, of BL Lacs with
those of samples of radio galaxies. This study revealed that BL Lacs
are found in regions characterized by a significantly lower galaxy
density with respect to FRIs, challenging predictions of the
unification scenario (e.g., \citealt{urry95}). They noted that instead
large-scale environments of BL Lacs are statistically consistent with
those of the FR~0s. This implies that  a substantial fraction of
BL~Lacs must be associated with FR0 sources oriented at a small
angle from the line of sight and that FR0s are able to produce
relativistic jets.

To test these suggestions VLBI observations of FR0s are necessary
to check the presence of small scale jets and to study their properties and
evolution.
\citet{cheng18} cross-matched the FR0CAT catalogue with
the mJy imaging VLBA exploration at 20 cm (mJIVE-20) catalog and the Astrogeo
database (see details and references in \citealt{cheng18}).
They found 14 FR0CAT sources with VLBI data, with fluxes at 1.4 GHz in the $\sim$50-400 mJy range. In the  VLBI images all of them show a radio structure. In four
sources the high brightness temperature suggests Doppler boosting; in most of the 
sources the core emission is likely blended with that of the inner jet.  They discuss that the  VLBI properties of these FR0s are consistent with low-power compact symmetric object (CSO) properties, but observations of more FR0s are necessary to investigate this point. The authors noted that these 14 sources are more compact and brighter than the
FR0 general population introducing possible selection effects. 

In a follow up project, \citet{chen21} selected, from the 14 sources studied by
\citet{cheng18}, the 8 FR0 that are resolved
sources on parsec scale and without multi-epoch VLBI observations. They obtained EVN and VLBA data to investigate the presence of proper motions, to identify the core position, and to classify the radio source structure. In all eight sources they confirmed the presence of compact structures.
In more detail: six sources
show a two-sided morphology and two are one-sided jet sources. The low proper
motion detected and the presence in most sources of a two-sided
structure suggest a relatively large viewing angle and/or a low bulk velocity
of FR0 jets.

\citet{baldi21c} presented e-MERLIN observations of five FR0s and preliminary results from 1.7 GHz EVN observations of ten FR0s, which are   discussed in
more details here. The main result is that the large fraction of two-sided structures further support the presence of
mildly relativistic jets in FR0s.

In this scenario we present here the results of EVN (at 1.7 GHz) and/or VLBA observations (at 1.5 and 5 GHz) of 18 FR0 sources. In Sect. 2 we  describe these observations and discuss the data reduction. The results are presented in Sect. 3.
In Sect. 4 we discuss FR0 parsec scale properties, and  compare our results with
the literature data. Our conclusions are presented in Sect. 5.

The intrinsic parameters quoted in this paper are computed assuming 
a $\Lambda$CDM cosmology with $H_0$ = 70 km s$^{-1}$Mpc$^{-1}$,
$\Omega_m$ = 0.3, and $\Omega_{\Lambda}$ = 0.7.

\section{Observations and data reduction}

\begin{figure}
   \centering
   \includegraphics[width=8cm]{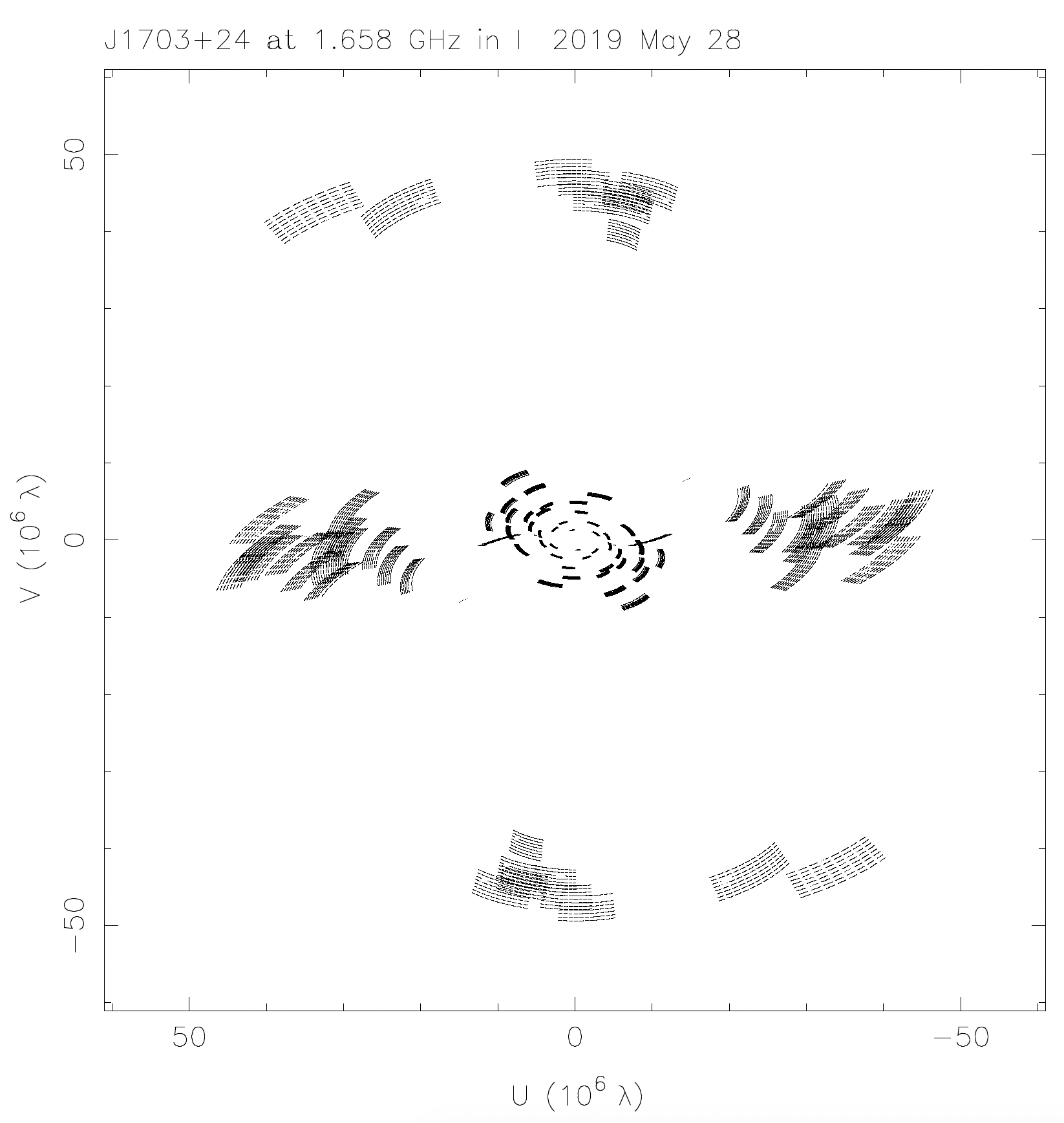}
      \caption{uv-coverage for  source J1703+24 observed with   EVN.
              }
         \label{Fig1}
   \end{figure}

Among the 25 FR0s studied with JVLA observations in the A array \citep{baldi15,baldi19}, we selected
sources with a core flux density at 1.4 GHz greater than 20 mJy and z $<$ 0.05.
The  11 sources selected were observed with EVN at 1.7 GHz. Observations
were carried in May 2019 in phase reference mode with a data rate of
1024 Mbit/sec (project EG100).
Nearby phase calibrators selected
from the VLBA calibrator list, and target sources were observed with a $\sim$ 5 min cycle. The total observing time was $\sim$ 2hr/source
with at least 60 min on the target source and the remaining time on calibrator
source and motion. Data were correlated at the JIVE correlator in Dwingeloo.
%
In Fig.~\ref{Fig1} we show the typical uv-coverage obtained using the following array:
Jodrell (Lovell), Westerbork single antenna, Effelsberg, Medicina, Onsala 25m,
Urumqi, Torun, Hartebeesthoek, Svetloe, Zelenchukaskaya, Badary, and Irbene.
Data calibration was carried out in the standard mode using the EVN pipeline.
Data reduction was done in AIPS, using the task IMAGR to obtain images.
Self-calibration in phase was applied (when possible because of the
low correlated flux density) to improve the quality of the images.
The final parameters are reported in the next section.

\begin{figure*}
   \centering
   \includegraphics[width=17.5cm, angle=0]{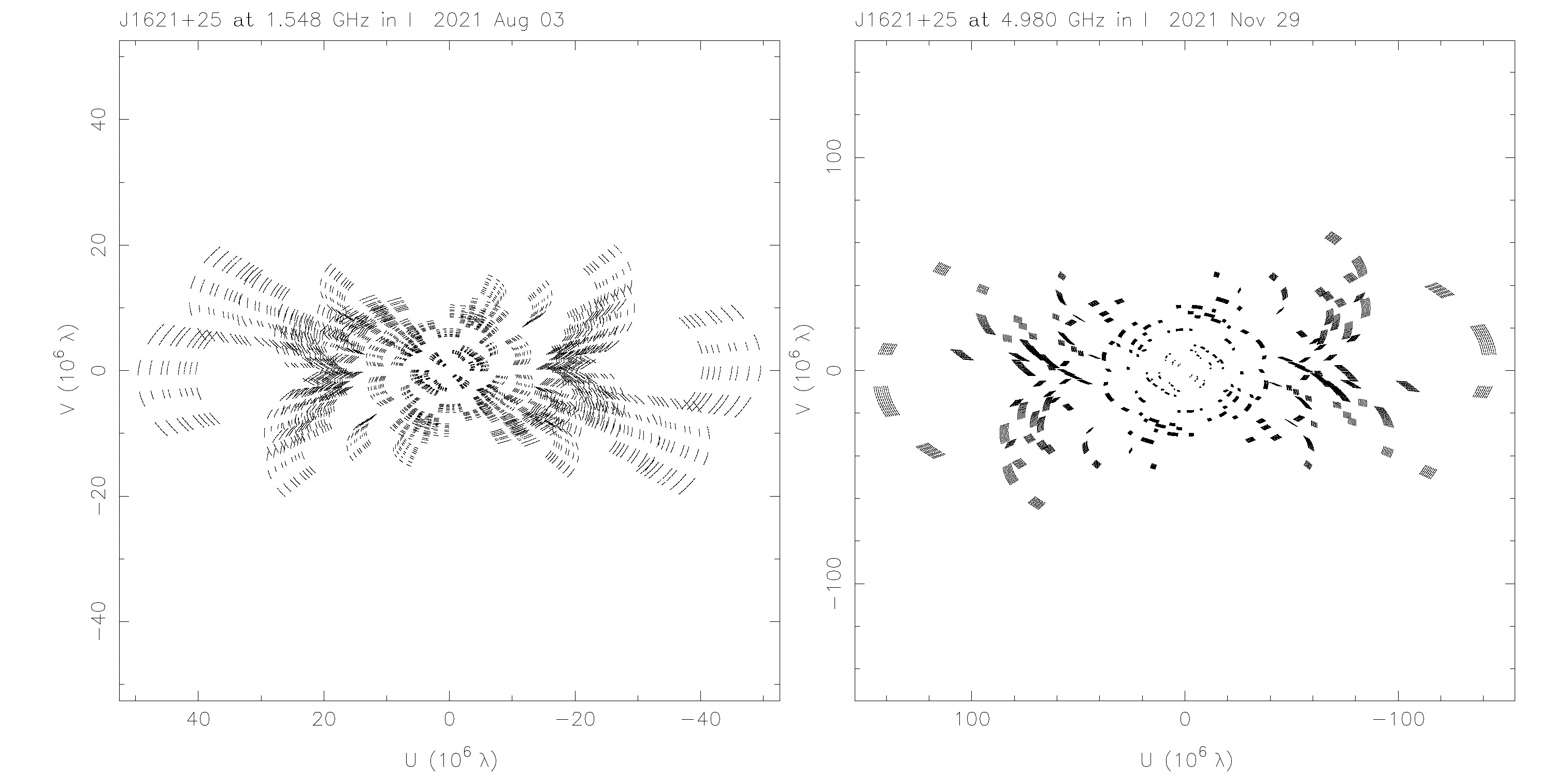}
   \caption{uv-coverage for  source J1621+25 observed with   VLBA at 1.5
   GHz (left) and at 5 GHz (right).}
   \label{Fig2}                                                   
    \end{figure*}

Moreover, in the time range September -- December 2021 we observed 11 FR0s with the VLBA
at 5 and 1.5 GHz (project BG272). Selected sources are characterized by a complex
and steep radio spectrum in the high-frequency range, with a flatter or
self-absorbed spectrum in the  low-frequency range. Sources were observed in
phase-reference mode at 5 and 1.5 GHz for 2 hr each including calibration and
dwell time. After the correlation at the NRAO Socorro correlator, data were
imported in AIPS and calibrated using the standard technique. Images were
produced using the AIPS task IMAGR; bad data were flagged mostly thanks to the
task TVFLG; self-calibration in phase only was
applied when possible to improve the image quality.
Dual frequency observations are useful to properly identify the
radio core and turbulence regions using spectral index information and
comparing different source morphology; 
we have high-resolution images of the source core and lower resolution, high-sensitivity images of the diffuse steep spectrum regions.
%
In Fig.~\ref{Fig2} the typical uv-coverage of a source at the two frequencies is shown.

 \section{Results}

In Table~\ref{tab1} we present the parameters of the images obtained from these observations and in
 Table~\ref{tab2} we present estimated source parameters from the images.
 The core structure was identified from spectral index information, when available. In other sources it was identified with the  brightest compact component. More details are reported in the next subsection (comments on the individual sources). Core parameters were measured using the task JMFITS in AIPS. This task fits up to four Gaussian components to a portion of an image. The starting model is a Gaussian with a width equal to the half power beam width of the image, centered on the peak of the component. In most sources a convergence is obtained after 40--50 iterations. In diffuse sources we integrated the flux density within the 2$\sigma$ area to avoid negative regions. We checked that the total flux density is not sensitive to a larger integration area. The source size is obtained from the deconvolved Gaussian fit in slightly extended regions, while it is directly measured on the image for well-resolved sources. 
 The size uncertainty is 0.1 times the HPBW for unresolved or slightly extended sources (Gaussian fit) and $\sim$ 0.3 times the HPBW in resolved sources. The source PA uncertainty is a few degrees.  We note that we produced images with a redundant cell size, approximately ten times smaller than the HPBW, for a better quality of the images.
 The flux density uncertainty was obtained combining the image noise level reported in Table~\ref{tab1} with the gain calibration uncertainty as    

$\Delta$ S = ((0.03$\times$S)$^2$ + noise$^2$)$^{0.5}$, \\
where S is the flux density; the calibration uncertainty has been evaluated at $\sim$ 3\%. We note that, given the low flux density of most of our sources, the density flux uncertainty is noise dominated. In a few sources the flux density uncertainty can be higher because of a peculiar source structure or other problems commented on in single source notes (Sect.~\ref{notes}).

\begin{table*}
\caption{Observational parameters}
\begin{footnotesize}
\begin{center}
\begin{tabular}{lcccccc}
\hline
\noalign{\smallskip}
Name        & EVN 1.7 & EVN 1.7  & VLBA 1.5 & VLBA 1.5 & VLBA 5 & VLBA 5 \\
            & HPBW    & noise    & HPBW     & noise    & HPBW   & noise   \\
            & mas     &$\mu$Jy/beam   & mas      &   $\mu$Jy/beam& mas    & $\mu$Jy/beam\\
\noalign{\smallskip}
\hline
\noalign{\smallskip}

J0020-00    & --      &  --      & 8 x 4    & 59       & 2.5 x 2.5  & 17    \\
J0115+00    & --      &  --      &10 x 4    & 46       & 3 x 2      & 18    \\
J0907+32    & 15 x 7  & 50       &10 x 10   & 77       & 4 x 3      & 17  \\
            &         &          &15 x 15   & 54       &            &     \\
J0930+34    & 16 x 9(-8)& 35       & --       & --       & --         & -- \\
J0943+36    & 17 x 7(-25)& 198      & --       & --       & --         & -- \\
J1025+10    &21 x 11(-5)& 100      & --       & --       & --         & -- \\
J1040+09    &20 x 20    & 120      & 14 x 13  & 15       & 5 x 5      & 20 \\
J1136+51    &  --     & --       & 13 x 8(-20)& 50       & 4 x 2.5    & 30 \\
J1213+50    & 25 x 25   & 270      & --       & --       & --         & --\\
J1230+47    & 24 x 15 & 140      & --       & --       & --         & --\\
J1530+27    &21 x 12(-10)&400      & --       & --       & --         & --\\
J1559+25    &15 x 10(-25)& 70      & --       & --       & --         & --\\
J1621+25    & --      & --       & 12 x 12  &	50     & 2.5 x 2.5  & 30 \\
J1628+25    &20 x 12(-40)& 100     & 4 x 3.5  & 35       & 0.9 x 0.9  & 17 \\
J1658+25    & --      & --       & 6 x 6    & 37       & 2.5 x 2.5  & 31 \\
J1703+24    &15 x 10(-14)& 50      & 5 x 3    & 29       & 3 x 2      & 25 \\
J2336+00    & --      & --       & 10 x 10  & 30       & 5 x 3      & 40 \\
J2346+00    & --      & --       & 5 x 5    & 28       & 2 x 2      & 15 \\
%
\noalign{\smallskip}
\hline
\noalign{\smallskip}
\label{tab1}
\end{tabular}
\end{center}
\end{footnotesize}
{Col. 1: source name; Col. 2: EVN HPBW at 1.7 GHz, the orientation angle is
 in parentheses, if not reported it is at zero degree or it is a circular beam; Col. 3: noise level (1 $\sigma$) of EVN images at 1.7 GHz; Col. 4: VLBA HPBW at 1.5 GHz, the orientation angle is in parentheses, if not reported it is at zero degree or it is a circular beam; Col. 5: noise level (1 $\sigma$) of VLBA images at 1.5 GHz; Col. 6: VLBA HPBW at 5 GHz, the orientation angle is
 in parentheses, if not reported it is at zero degree or it is a circular beam; Col. 7: noise level (1 $\sigma$) of VLBA images at 5 GHz.} 
\end{table*}

 Images of resolved sources are in Figures~\ref{Fig3} to \ref{Fig15}. No image is presented for unresolved sources since they appear as a beam image and no information (e.g., field sources or noise structures) is present. Single sources are discussed in Sect.~\ref{notes} where more details are presented.
 
\begin{table*}
\caption{Source data}
\begin{footnotesize}
\begin{center}
\begin{tabular}{lccrrrrrl}
\hline
\noalign{\smallskip}
Name & RA &  z    &EVN 1.7&VLBA 1.5&VLBA 5& LAS & PA  & Note \\
     &DEC &       & core(mJy)& core(mJy)& core(mJy) & mas & deg &      \\
     &    &       & total(mJy)&total(mJy)&total(mJy)& pc  &     &      \\
\noalign{\smallskip}
\hline
\noalign{\smallskip}

J0020-00& 00 20 34.700& 0.0720 &  -- & 0.94(0.07)& 0.35(0.02) & 10 & 160  & one-sided \\
        &-00 28 14.40 &        &  -- & 1.48(0.08)& 0.82(0.03) & 14 & 140  &           \\
J0115+00& 01 15 15.702& 0.0450 &  -- & 7.26(0.05)& 5.77(0.02) & 40 & 95   & two-sided \\
        & 00 12 48.59 &        &  -- &38.42(1.10)&35.34(1.00) & 35 &      &           \\
J0907+32& 09 07 34.827& 0.0490 &  -- & 0.45(0.05)&$<$0.04&65  & 80 &double+core \\
        & 32 57 22.80 &        & 13.28(0.70)&12.47(0.07)& ND  & 62 &     &Spiral galaxy\\
J0930+34& 09 30 03.560& 0.0420 & --  & --        & --   & 15  & 90 & double   \\
        & 34 13 25.34 &        & 7.18(0.36) & -- & --   &12  &      & structure \\
J0943+36& 09 43 19.159& 0.0223 & 140(7.0)& --      & --   & P  & --   &unresolved \\
        & 36 14 52.07 &        &   --  &   --    &  --  & -- &      &variable    \\
J1025+10& 10 25 44.213& 0.0457 & 106(5.0)& --      & --   & P  &110  &see text   \\
        & 10 22 30.48 &        &   --  &   --    &  --  & -- &     &     \\
J1040+09& 10 40 28.373& 0.0195 & --    & 2.45(0.20)&3.08(0.25) & 90 & 180  & two-sided \\
        & 09 10 57.25 &        & 14.57(0.74)&19.25(0.58)& 5.68(0.28) & 36 & & complex  \\
J1136+51& 11 36 37.148& 0.0498 & --  & 1.43(0.05) & 0.30(0.03) & P  &  --  &unresolved \\
        & 51 00 08.44 &        & --    &   --    &  --  & -- &      &steep spectrum  \\
J1213+50& 12 13 29.280& 0.0308 & --    &   --    & --   & 15 & 145  &see text  \\
        & 50 44 29.38 &        & 54.7(2.8) &   --&  --  &  9 &      &       \\  
J1230+47& 12 30 11.852& 0.0391 &11.85(0.83)&   --    & --   & 100&  0   &two-sided \\
        & 47 00 22.63 &        & 68.8(2.1) &   --    &  --  & 77 &      & symmetric \\
J1530+27& 15 30 16.153& 0.0325 & 48.5(2.4) & --      & --   & 15 &  65  &one-sided? \\
        & 27 05 51.01 &        & 62.6(3.1) &   --    &  --  & 10 &      &        \\
J1559+25& 15 59 51.615& 0.0447 & 6.98(0.71)  & --      & --   & P  & --   &see text \\
        & 25 56 26.33 &        & --    &   --    &  --  & -- &      &      \\
J1621+25& 16 21 46.069& 0.0480 & --    & 0.50(0.06)& 0.28(0.03) &180 & 90 &two-sided  \\
        & 25 49 14.45 &        & --    & 5.95(0.19)& 0.30(0.03) &169 & 130& z-shaped  \\
J1628+25& 16 28 46.129& 0.0401 & --    & 8.27(0.31)& 7.94(0.28) &12 &115-90& 2-sided  \\
        & 25 29 40.97 &        & 27.4(1.4) & 27.62(0.83)&26.50(0.80)&10  &      &   \\
J1658+25& 16 58 30.054& 0.0327 & --   & 6.51(0.10)& 10.26(0.16)&73  &  60  & 2-sided  \\
        & 25 23 24.97 &        & --    & 13.83(0.17)&12.94(0.16) &48  & &distorted  \\
J1703+24& 17 03 58.503& 0.0310 & 4.00(0.20)  & ??      & ??   &40  & 70?  & core?  \\
        & 24 10 39.52 &        &11.00(0.55)& 9.34(0.30)& 3.46(0.15) &25  &  & complex\\
J2336+00&             & 0.0763 & --    &  ND     &  ND  &--  & --   & not detected \\
        &             &        & --    &  --     &  --  &--  &      &   \\
J2346+00& 23 46 09.103& 0.0927 &  --   & 4.61(0.25)& 2.00(0.21) & 30 &105   & 2-sided \\
        & 00 59 08.83 &        & --    & 23.91(0.72)& 9.84(0.30) & 52 &      & z-shaped \\
\noalign{\smallskip}
\hline
\noalign{\smallskip}
\label{tab2}
\end{tabular}
\end{center}
\end{footnotesize}
{
Col. 1: source name;
Col. 2: core position obtained with a gaussian fit;
Col. 3: source redshift;
Cols. 4,5,6: core (top) and total (bottom) flux density in mJy in the different available images;
Col. 7: largest measured size of the source in mas (top) and parsec (bottom);
Col. 8: source orientation in the image in degrees;
Col. 9: comments.
}
\end{table*}

Table~\ref{tab3} provides the source parameters of our sample
The jet-to-counter-jet ratio, reported in Col. 5 is the ratio of the surface brightness of the two opposite jets in the region near the VLBI core, $R_{\rm JC} =F_{\rm jet}/F_{\rm cjet}$. When the counter-jet was not detected we report an upper limit obtained using the one-sigma noise level as counter-jet surface brightness. In other sources the ratio uncertainty is in parentheses; it does not affect observed beaming effects.

\begin{table*}
  \caption{Source properties}             
  \begin{footnotesize}
    \begin{center}
\label{tab3}      
\begin{tabular}{lcccc}  
  \hline\hline
  \noalign{\smallskip}
Name   & Log P$_{tot}$ ($\nu$) & Log P$_{core}$ ($\nu$) & S$_{VLBI}$/S$_{core-JVLA}$  & jet-to-cjet ratio \\
          &      W/Hz    &   W/Hz      &       &         \\      
\hline                        
J0020-00  & 22.75 (1.5)  & 21.65 (5.0) &  1.59 & $>$ 2.7 \\ 
J0115+00  & 23.26 (1.5)  & 22.44 (5.0) &  1.05 & 1 (0.1) \\
J0907+32  & 22.88 (1.7)  & 21.41 (1.5) &  3.34 & 1 (0.2) \\
J0930+34  & 22.47 (1.7)  &    --       &  0.57 & $>$ 10   \\
J0943+36  &    --        & 23.20 (1.7)?&  0.50 &  --   \\
J1025+10  & 23.72 (1.7)  &     --      &  0.92 & --     \\
J1040+09  & 22.22 (1.5)  & 21.42 (5.0) &  0.77 & 1 (0.2) \\
J1136+51  & 21.92 (1.5)  & 21.25 (5.0) &  1.23 &        \\ 
J1213+50  & 23.08 (1.7)  &             &  0.74 &  1 (0.1 \\
J1230+47  & 23.39 (1.7)  & 22.63 (1.7) &  1.35 & 1.4 (0.1)  \\
J1530+27  & 23.18 (1.7)  & 23.07 (1.7) &  1.39 & $>$ 8  \\
J1559+25  &              & 22.57 (1.7) &  0.52 &        \\
J1621-25  & 22.51 (1.5)  & 21.18 (5.0) &  2.70 & 1 (0.1) \\
J1628+25  & 23.02 (1.5)  & 22.47 (5.0) &  1.14 & 1 (0.1) \\
J1658+25  & 22.53 (1.5)  & 22.40 (5.0) &  1.01 & 1 (0.1) \\
J1703+24  & 22.39 (1.7)  &   ???       &  0.44 & 1 (0.1) \\
J2236+00  &  ---         &   ---       &  ...  &          \\
J2346+00  & 23.71 (1.5)  & 22.64 (5.0) &  2.63 & 1 (0.1)  \\

\hline                                   
\end{tabular}
\end{center}
\end{footnotesize}
{Col. 1: source name;
Col. 2: total radio power in W/Hz, in parentheses the observed radio frequency in GHz;
Col. 3: core radio power in W/Hz, in parentheses the observed radio frequency in GHz;
Col. 4: the ratio of the correlated VLBI flux density to the core flux density in JVLA observations;
Col. 5: the jet-to-counter-jet surface brightness ratio in the region near to the VLBI core, the ratio uncertainty is in parentheses.}
\end{table*}

 \subsection{Comments on the individual sources}
\label{notes}

\begin{figure}
   \centering
   \includegraphics[width=7cm]{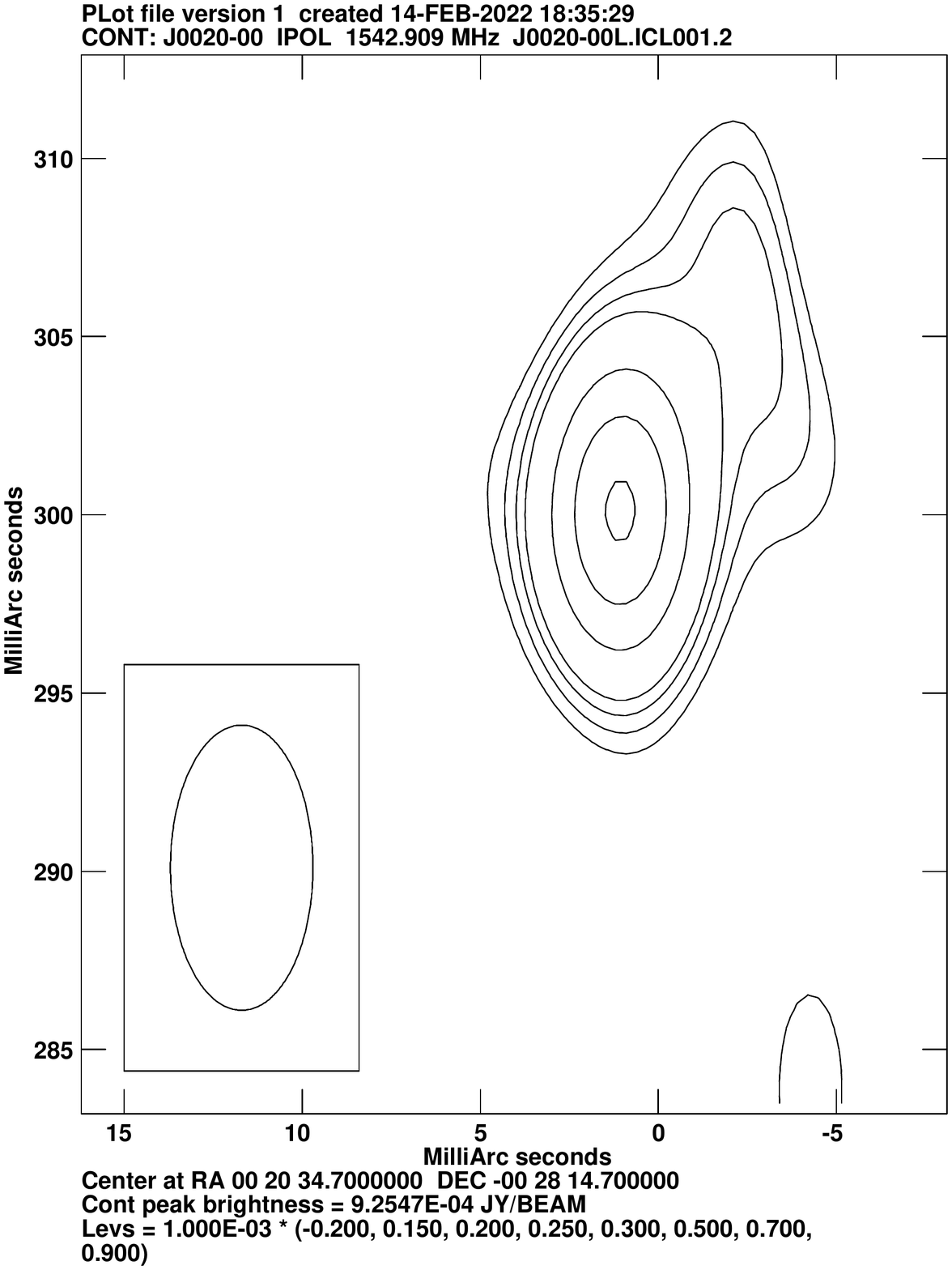}                                        
   \caption{Contour image of J0020-00 obtained with   VLBA at 1.5 GHz
   }
         \label{Fig3}
 \end{figure}

 {\bf J0020-00} - This source shows a two-sided symmetric structure in the
 JVLA image at 7.5 GHz published in \citet{baldi15}. eMERLIN show a core with scattered components along the  JVLA radio axis \citet{baldi21c}. In the VLBA images it shows (Fig.~\ref{Fig3}) a one-sided structure at 1.5 GHz at PA = 140$^\circ$ (the same as the arcsec structure)
and an inner core in the VLBA image at  5 GHz with a symmetric short extension along PA 160$^\circ$. The jet-to-counter-jet ratio at 1.5 GHz is $>$2.7. The JVLA core flux density at  7.5 GHz is significantly lower than the correlated flux density at 1.5 GHz
 (the ratio is 1.59), suggesting variability and/or complex nuclear spectral index.

\begin{figure}
   \centering
   \includegraphics[width=7cm]{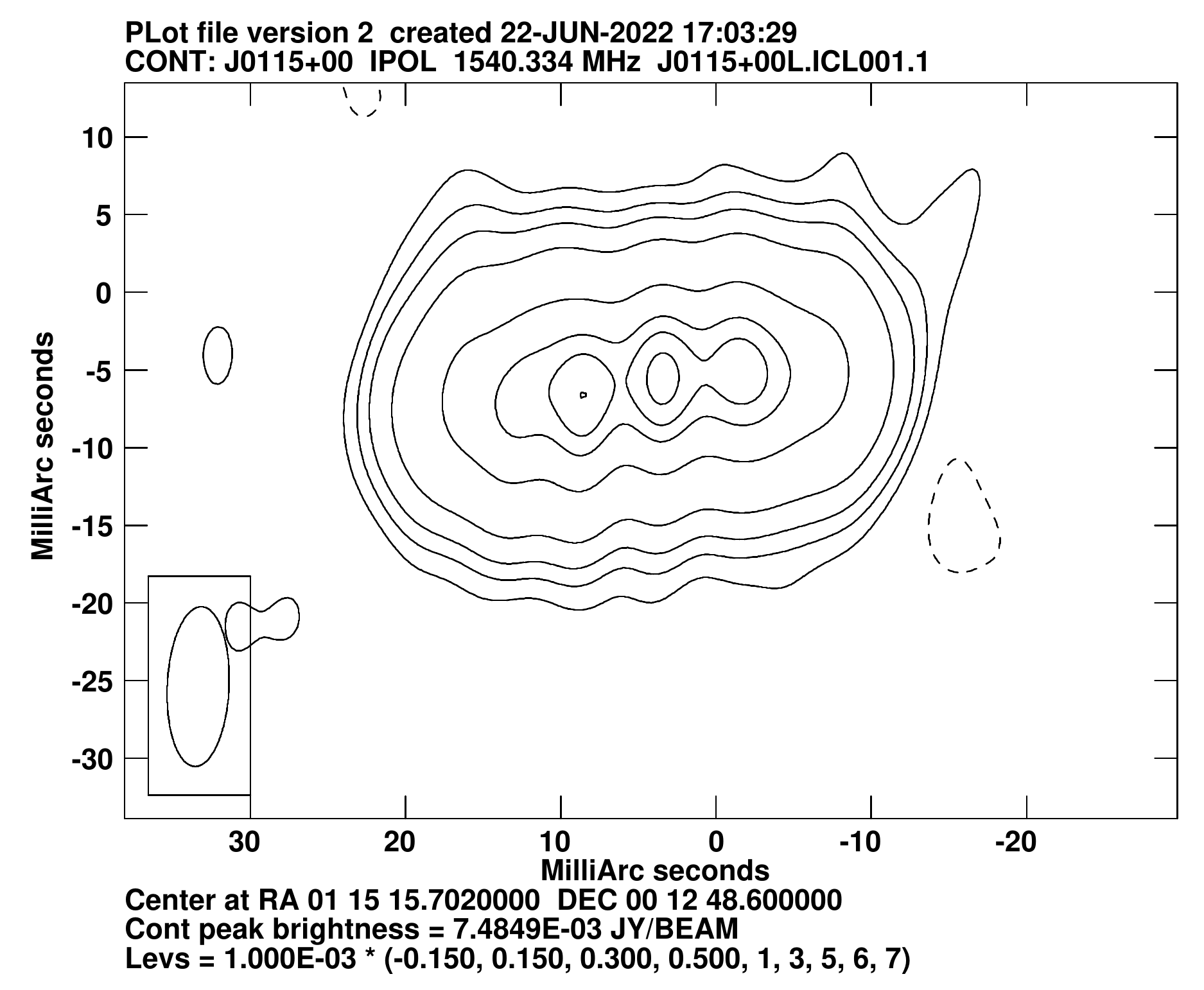}                                        
      \caption{Contour image of J0115+00 obtained with   VLBA at 1.5 GHz
              }
         \label{Fig4}
   \end{figure}

 {\bf J0115+00} - This source, unresolved in JVLA images \citep{baldi15},
 shows a symmetric two-sided structure in VLBA images at 1.5 (Fig.~\ref{Fig4}) and 5 GHz. The
 ratio of the VLBA correlated flux
 at 1.5 GHz to the JVLA flux density at 7.5 GHz
 is 1.05, suggesting that in the VLBA image all the nuclear structure
 is visible. The jet-to-counter-jet ratio is $\sim$ 1. The source spectrum in both the JVLA and VLBA
 images is flat.

\begin{figure}
   \centering
   \includegraphics[width=7cm]{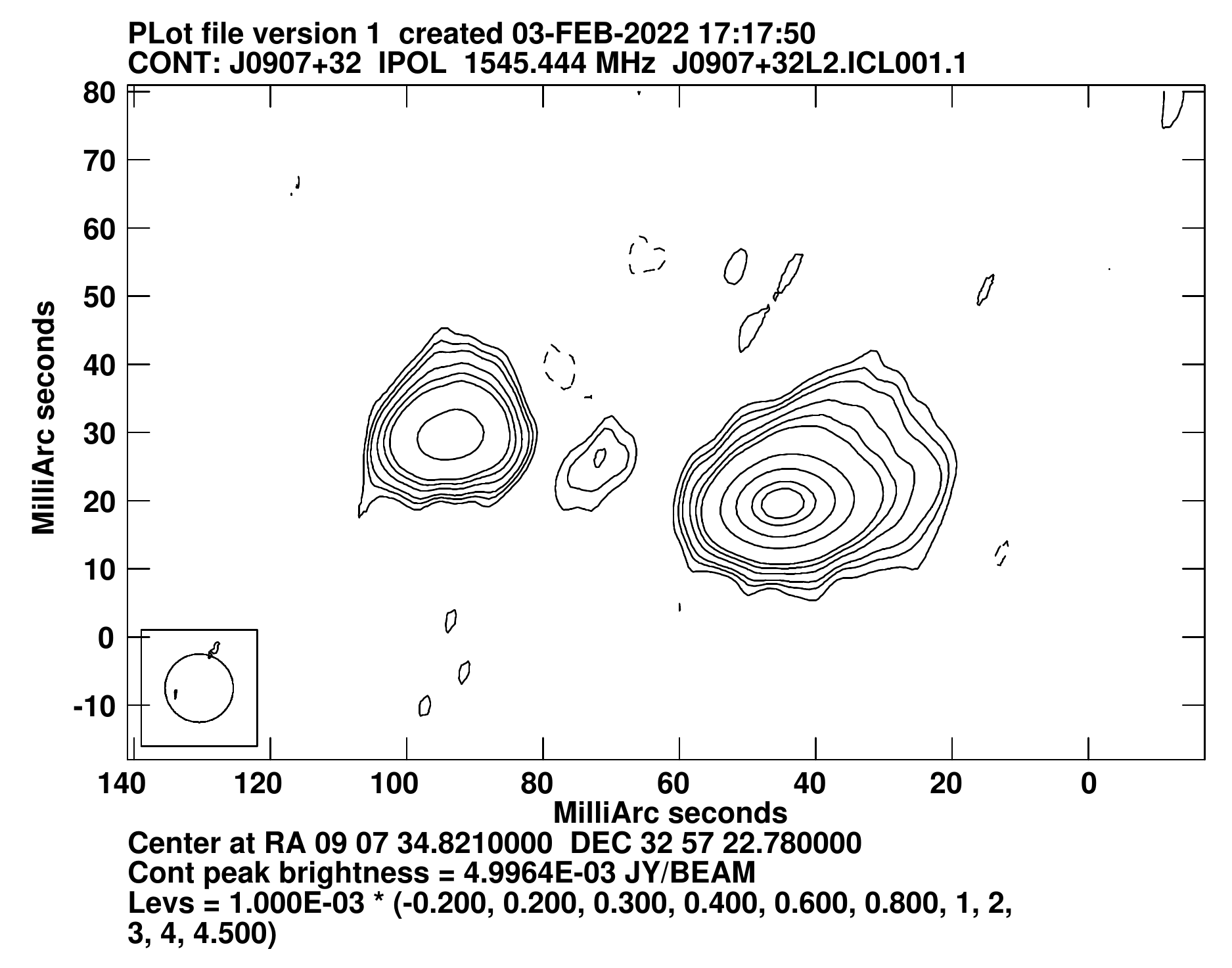}                                        
      \caption{Contour image obtained with   VLBA at 1.5 GHz of J0907+32
              }
         \label{Fig5}
 \end{figure}

 {\bf J0907+32} - This peculiar source is identified with a spiral galaxy
 and should not be included among the  FR0 sources. However, because of its peculiar
 structure we report here data and images. We do not use it in statistical
 considerations. A similar source (B2 0722+30) was found by \citet{liuzzo13}, who studied a sample of compact FRI sources.
 In JVLA images \citep{baldi19} J0907+32 shows a nuclear emission and a symmetric
 structure identified with the disc emission easily confused with two-sided
 jets. Interestingly the nuclear emission is slightly resolved in the direction perpendicular to the disk. EVN images at 1.7 GHz and VLBA at 1.5 GHz shows
 two symmetric small-scale lobes in agreement with the nuclear arcsecond
 extension perpendicular to the disk. In the VLBA image a possible central emission is present (Fig.~\ref{Fig5}). The VLBA data at 5 GHz  completely resolve this structure and no significant structure is
 visible. It is a puzzling steep spectrum source, suggesting  AGN activity in
 the core of this spiral galaxy but with no jet-like structure. The presence of a similar structure in B2 0722+30, as noted before, could suggest the possibility of FR0-like structures in a few cores of spiral galaxies. More data are necessary to discuss this point.

\begin{figure}
   \centering
   \includegraphics[width=7cm]{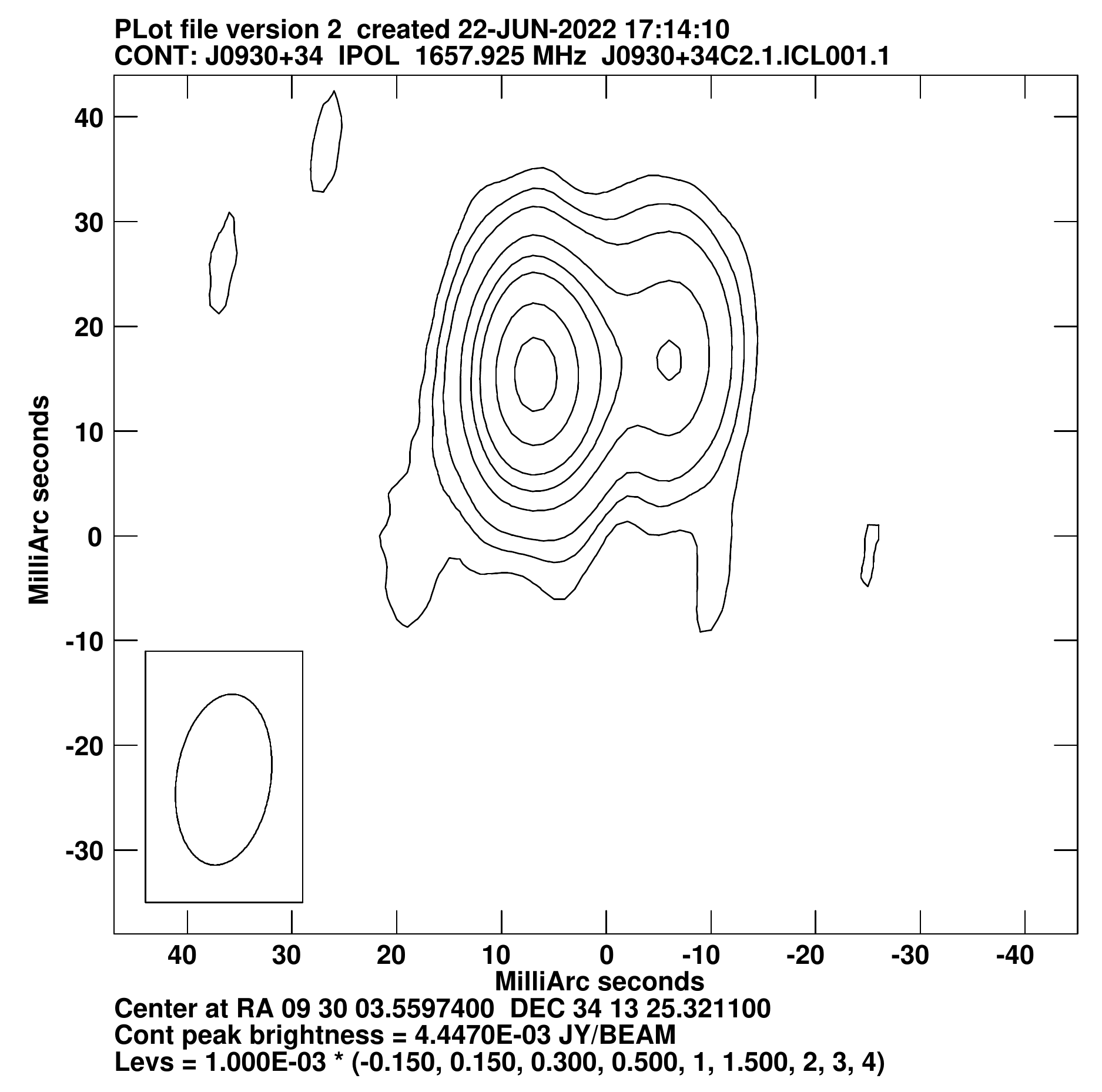}                                        
      \caption{Contour image of J0930+34 obtained with   EVN at 1.7 GHz
              }
         \label{Fig6}
 \end{figure}

 {\bf J0930+34} - This source, unresolved but with a steep spectrum in JVLA observations
  \citep{baldi19}, is resolved in EVN observations in a double structure (Fig.~\ref{Fig6}).
 The core
 identification is not clear; we note that the correlated
 flux density is significantly lower than the JVLA flux density at 7.5 GHz. The
 ratio is 0.57, due to the source steep spectrum and a possibly missing structure in
 VLBI image. In any case, if the stronger lobe (5.42 mJy) is the core and
 the secondary lobe (1.76 mJy) a one-sided jet, the jet-to-counter-jet ratio does not require a
 fast jet (v $>$ 0.4c).

  {\bf J0943+36} - This source is unresolved in our EVN observations and in VLBA
 observations at 8 GHz  \citep{cheng18}. The source flux density is 140 mJy and in the image no dynamic range problem is present. The unresolved source is the only visible structure, also in large field images. 
 The radio spectrum is inverted
 and it shows evidence of variability. This property and the relatively high
 radio power suggest a relativistic source oriented at a small angle from the line of sight.
 The ratio of the correlated EVN flux
 density to the JVLA flux density at 7.5 MHz {\bf \citep{baldi19}} is 0.50. This can be due to an
 inverted spectrum and source variability.

 {\bf J1025+10} - Unresolved in EVN observations at 1.7 GHz, it is resolved with
 a two-sided structure (13 pc in size) in the EVN 5 GHz image by \citet{cheng21}.
 In their image the inner core is extended at PA = 90$^\circ$
 and the two-sided jets are at PA 130$^\circ$,
 suggesting a Z-shaped structure in the inner ($\sim$ 1 pc) region. The source size in the 5 GHz image is in agreement with the unresolved structure at 1.7 GHz.
 In our image no dynamic range problem is present. The unresolved source is the only visible structure, also in large field images.
 The core flux density  ratio of the EVN correlated flux to the JVLA flux density at 7.5 GHz {\citep{baldi19}} is 0.92.
 The radio spectrum from the total flux density estimated with JVLA is peaked at 5 GHz in agreement with the source's small size.

\begin{figure}
   \centering
   \includegraphics[width=7cm]{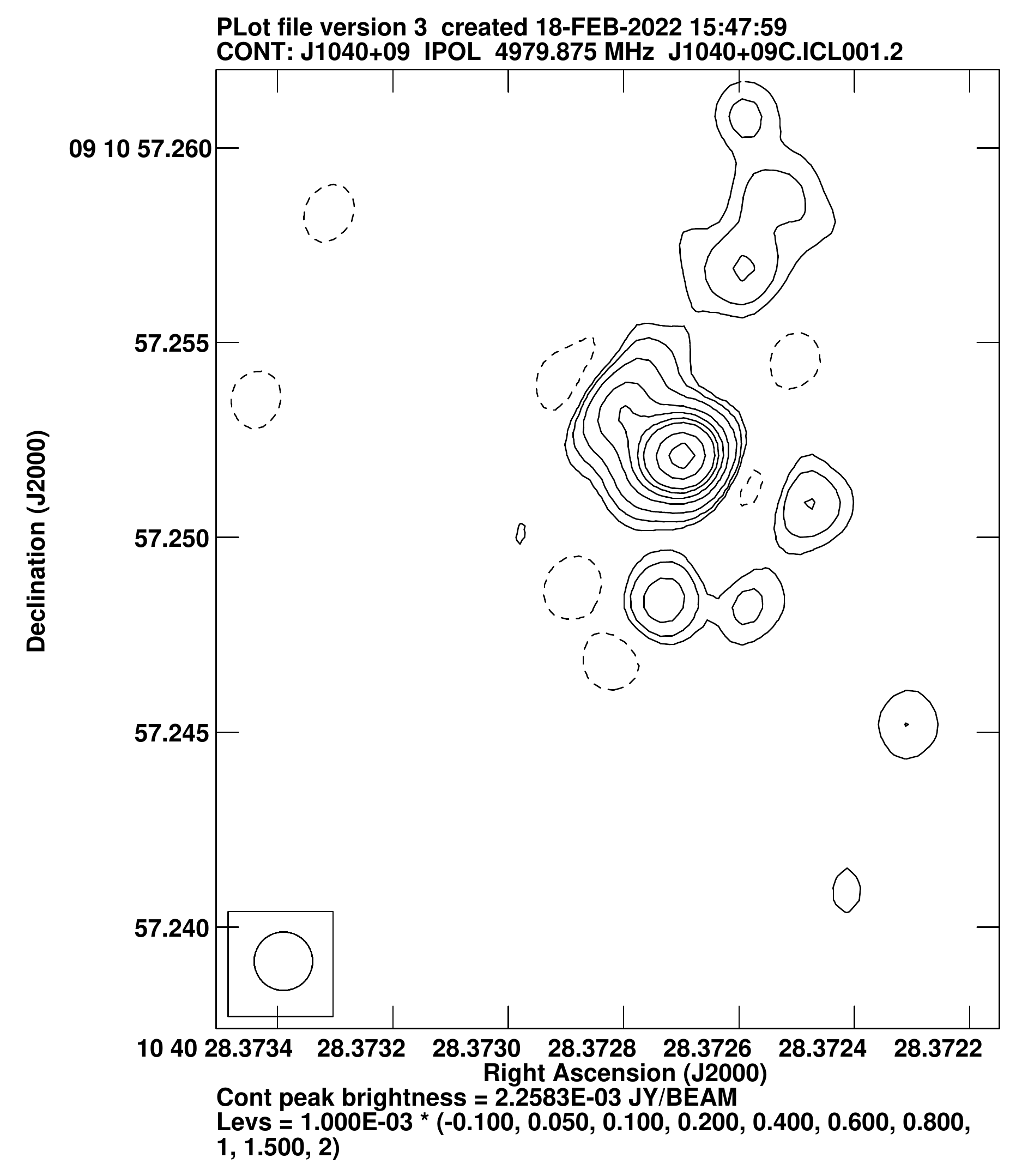}
      \includegraphics[width=7cm]{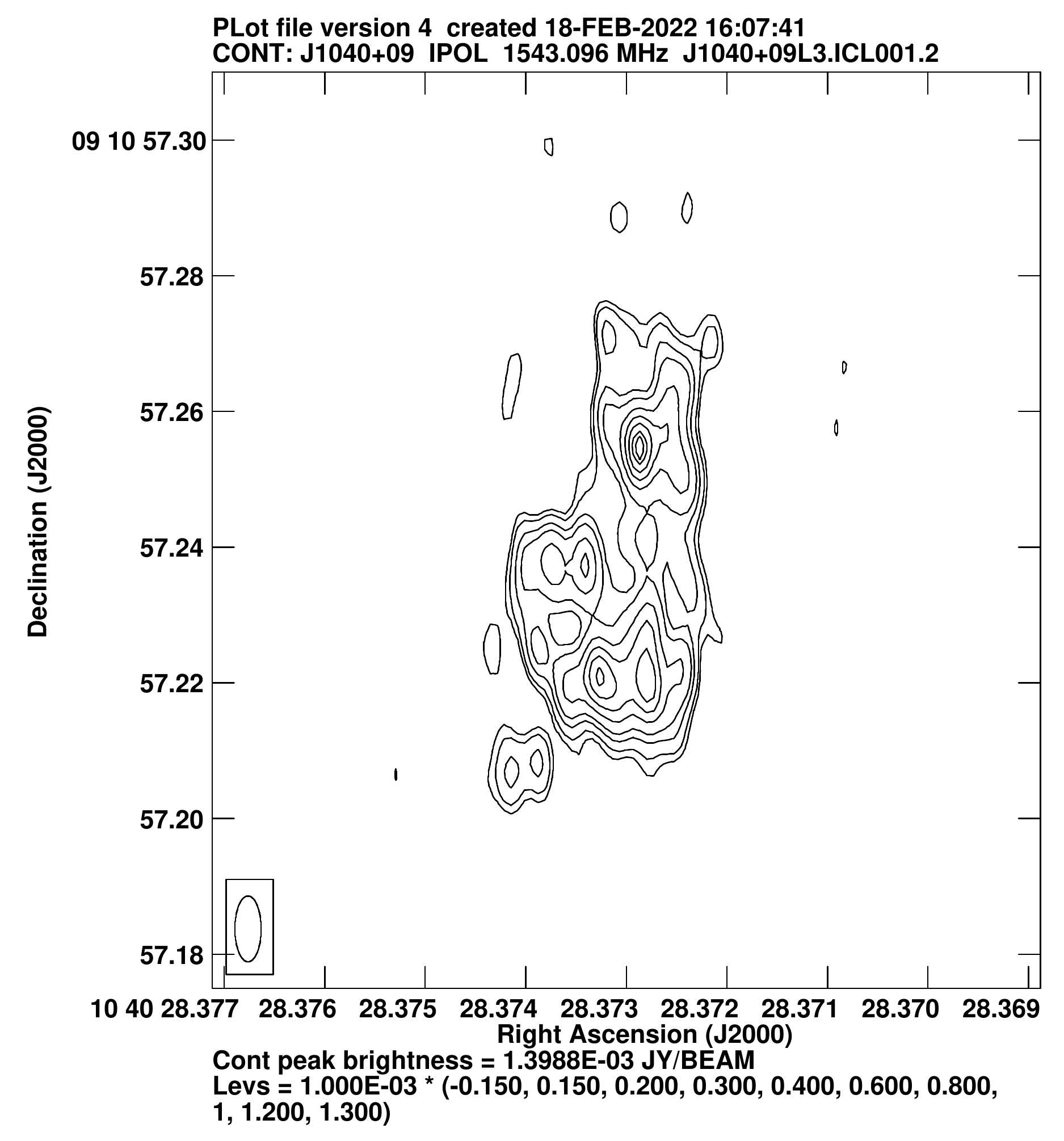}                                        
      \caption{Contour image of J1040+09 obtained with   VLBA at 5 GHz (top) and at
        1.5 GHz (bottom)
              }
         \label{Fig7}
 \end{figure}

 {\bf J1040+09} - This is a very complex source. The VLBA image at 5 GHz shows an unresolved core,
 a northern curved jet, and some structure in the south (Fig.~\ref{Fig7}, top panel).
 At 1.5 GHz the core is surrounded by a diffuse emission; the northern jet is extended
 $\sim$ 20 mas, transversely resolved, while the southern jet is extended $\sim$ 60 mas and it is heavily transversely resolved (Fig.~\ref{Fig7}, bottom panel).
 The classification of the structure as a jet is uncertain; it could be a parsec-scale lobe, given its size of 
 15.80 x 8.7 pc (PA 180$^\circ$). Lower resolution
 images confirm this structure, with a slightly larger size.
 The comparison between the correlated VLBI flux density
 and the flux density of the unresolved source imaged by JVLA at 7.5 GHz \citep{baldi19} is 0.77.

 Because of the low correlated flux, no self-calibration was possible. We carefully edited the data and compared the  1.5 and 5 GHz images with the same uv-range and angular resolution  between the images shown here. In these images the radio structures are in agreement; therefore, we are confident that the observed complex structure is real.

 {\bf J1136+51} - This is an unresolved source with a steep spectrum, $\alpha_{1.5}^5$ = 1.3,
 which could be an old source. No other radio source is visible, even  in large field images. The ratio of the VLBI correlated flux to the JVLA flux density \citep{baldi19} is 1.22.

\begin{figure}
   \centering
   \includegraphics[width=7cm]{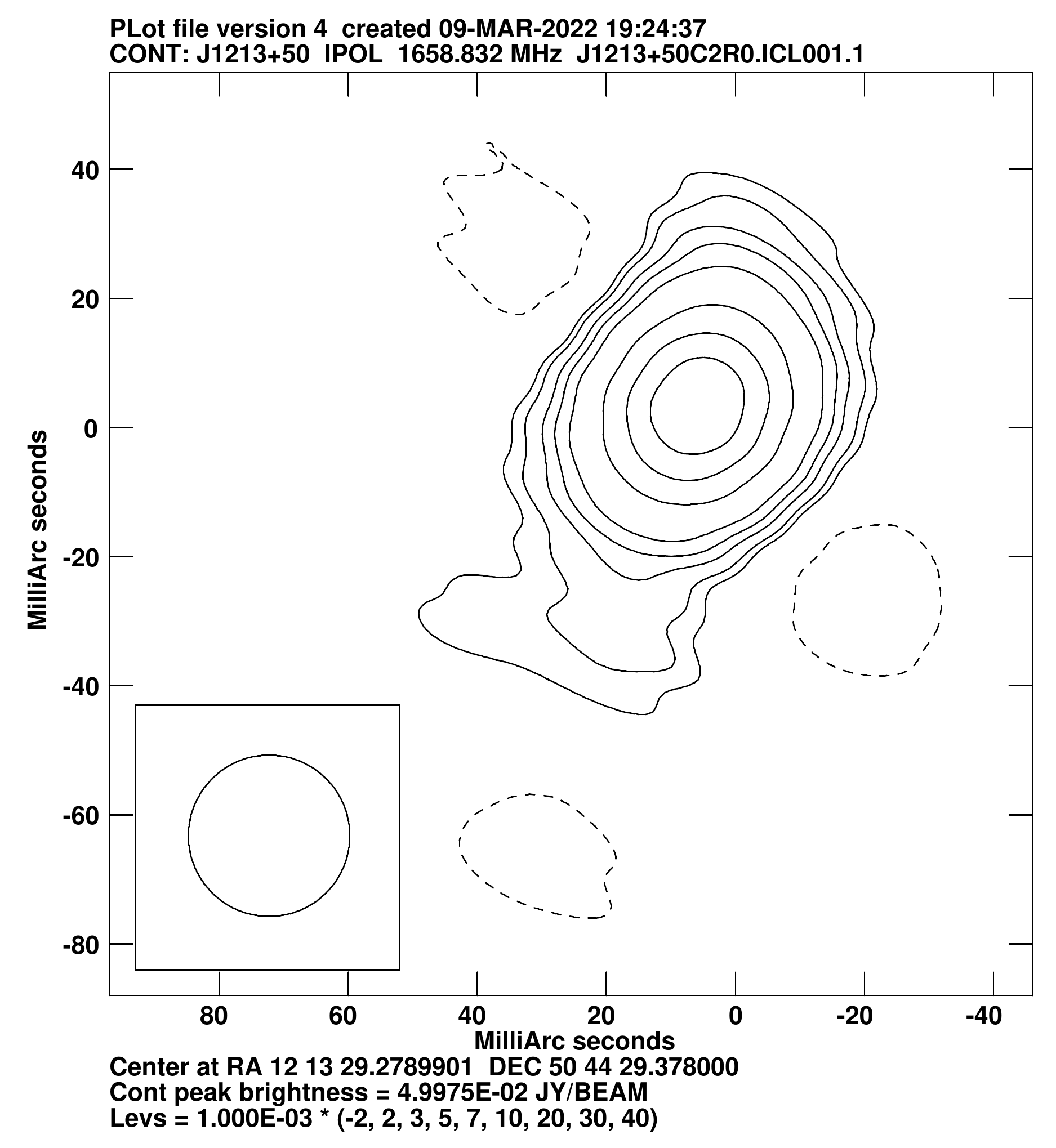}                                        
      \caption{Contour image of J1213+50 obtained with   EVN at 1.7 GHz
              }
         \label{Fig8}
 \end{figure}

 {\bf J1213+50} - This is a slightly extended source with a nearly symmetric structure in the
 EVN image at 1.7 GHz (Fig.~\ref{Fig8}). The source extension is at PA 145$^\circ$, but at PA 0$^\circ$ in the  JVLA 5 GHz image \citep{baldi19} and 90$^\circ$ in the  VLBA 8 GHz image by \citet{cheng18}. No jet boosting is evident in all images. The core flux density ratio is 0.74.

\begin{figure}
   \centering
   \includegraphics[width=7cm]{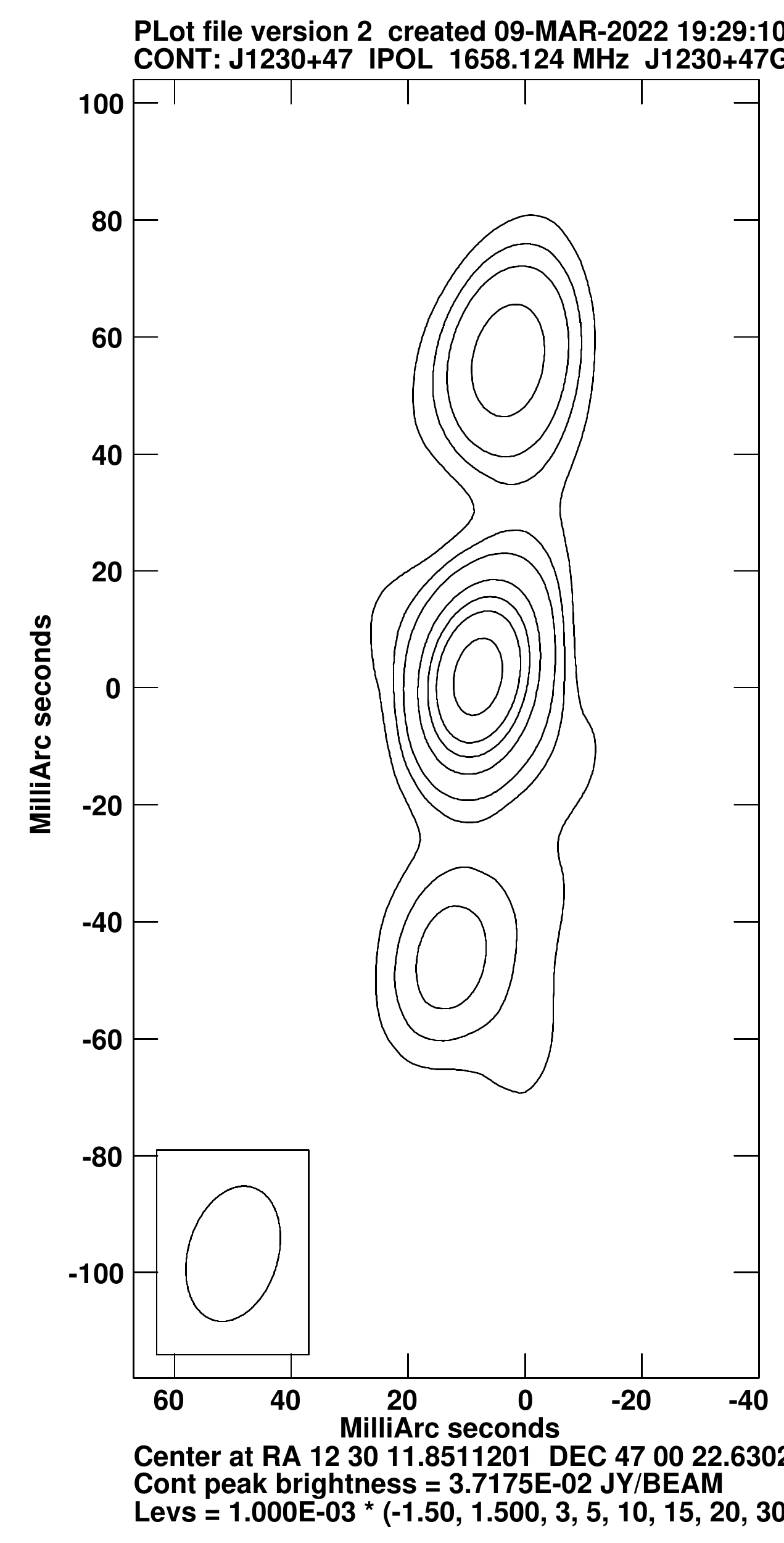}                                        
      \caption{Contour image of J1230+47 obtained with   EVN at 1.7 GHz
              }
         \label{Fig9}
   \end{figure}

 {\bf J1230+47} - The extended symmetric structure in the  EVN images at 1.7 GHz 
 is more than 70 pc in size (Fig.~\ref{Fig9}). The central core is resolved in the images by \citet{cheng18} in a two-sided structure 7-8 pc in size at the same
 PA of the more extended structure, suggesting a possible recurrent activity. The source is unresolved in JVLA images by \citet{baldi19}.

\begin{figure}
   \centering
   \includegraphics[width=7cm]{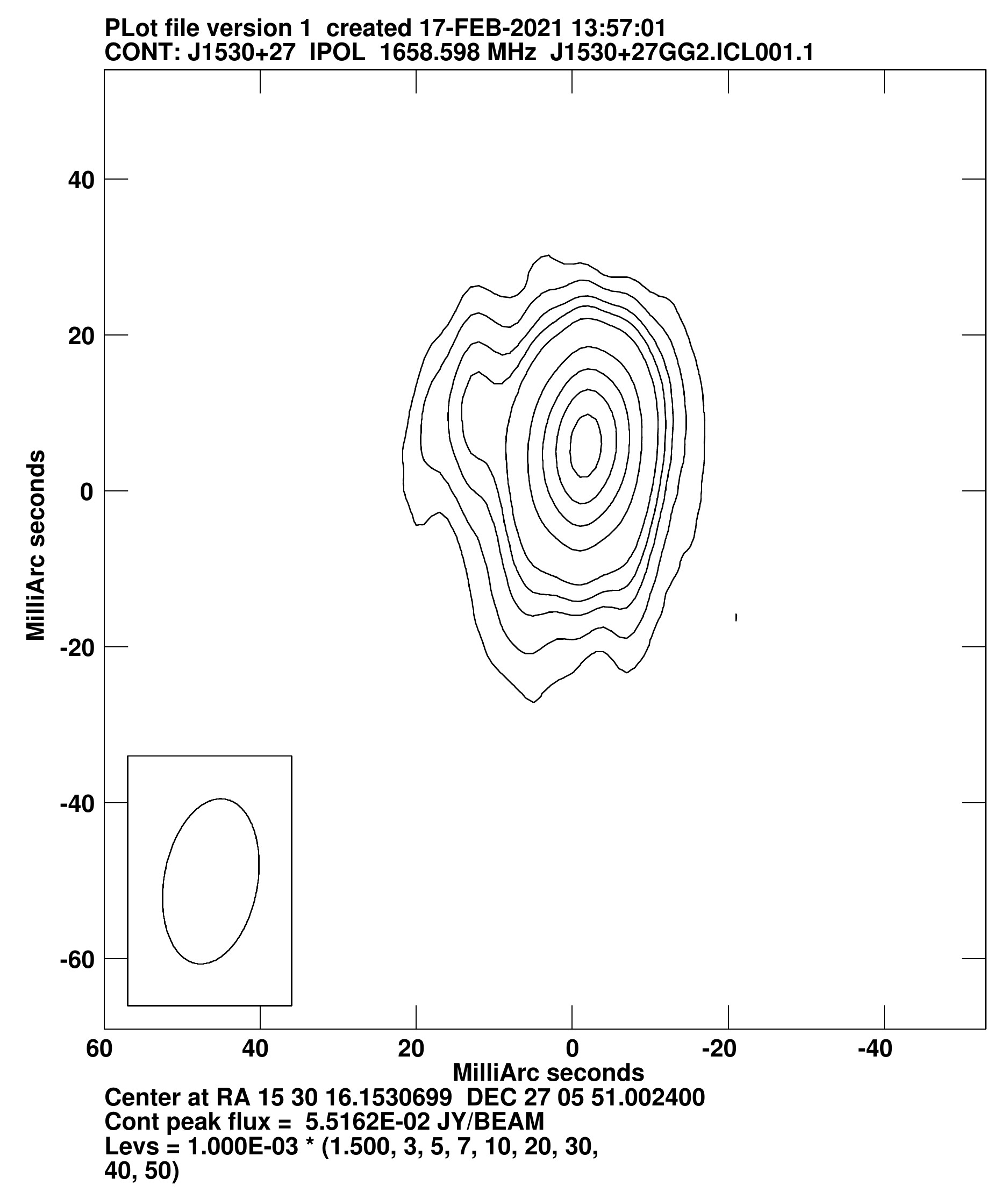}                                        
      \caption{Contour image of J1530+27 obtained with   EVN at 1.7 GHz
              }
         \label{Fig10}
 \end{figure}

 {\bf J1530+27} - A core dominates in the EVN images at 1.7 GHz with a short one-sided emission (Fig.~\ref{Fig10})
 possibly identified with a jet. Because of the low brightness of this structure, 
 Doppler boosting is not evident. The JVLA flux densities show an unresolved
 self-absorbed structure with the highest flux density at 5 GHz \citep{baldi19}.

 {\bf J1559+25} - This source is one-sided in JVLA images \citep{baldi19} with a jet size of  3 kpc;
 it is unresolved in our EVN data, but one-sided
 in the 8GHz VLBA image by \citet{cheng18}. The parsec- and kiloparsec-scale structure are at the same PA. The jet is
 transversely resolved in the VLBA image. The jet-to-counter-jet ratio in the VLBA image is higher than 6. The minimum velocity is 0.34c and the maximum orientation
 angle is 70$^\circ$. 

 Our image is not affected by dynamic range problems. No other radio structure is visible, even in large field images.

\begin{figure}
   \centering
   \includegraphics[width=7cm]{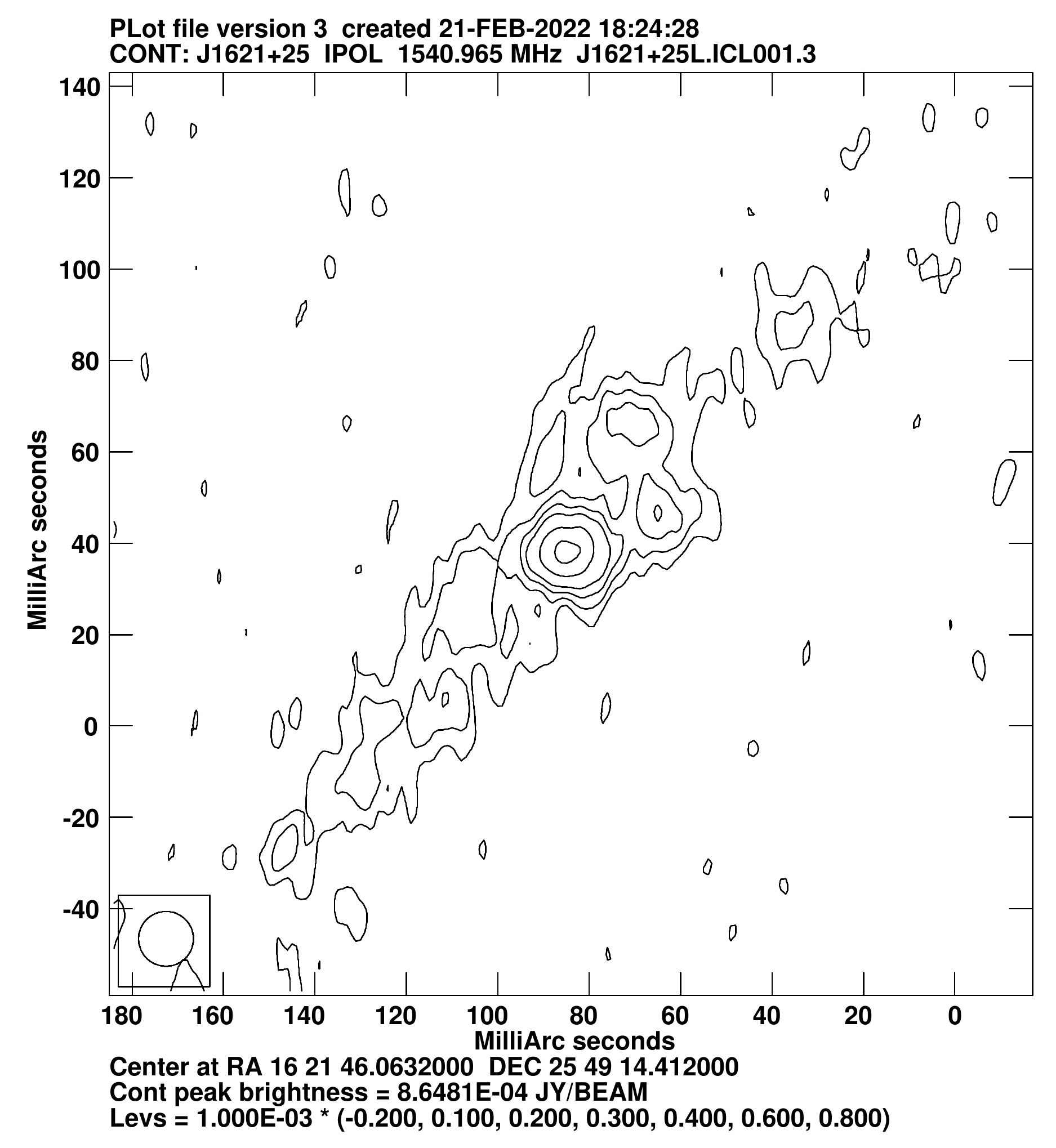}
      \includegraphics[width=7cm]{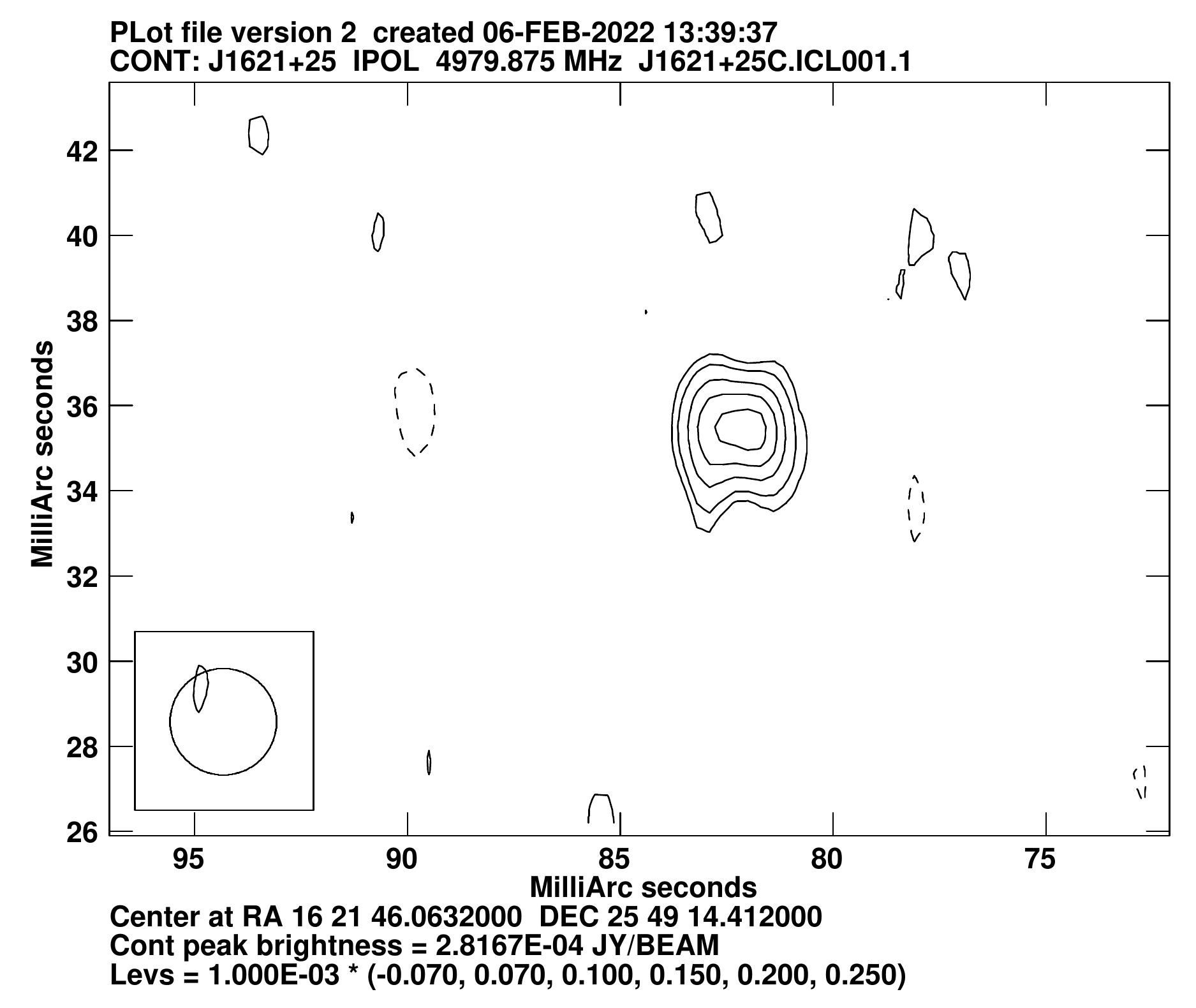}                                        
      \caption{Contour image of J1621+25 obtained with  VLBA at 1.5 GHz (top) and 5 GHz (bottom)
              }
         \label{Fig11}
     \end{figure}

 {\bf J1621+25} - In the 5 GHz and 1.5 GHz high-resolution VLBA images this source shows  a
 short (28 pc in size) symmetric nuclear structure at PA = 90$^\circ$.
 At lower resolution
 (HPBW = 12 x 12 mas),  a symmetric complex double-jet structure $\sim$ 170 pc in
 size   is clearly evident at PA 130$^\circ$. The jet is transversely resolved and its extension is noise limited (see Fig.~\ref{Fig11}).
 The correlated flux density is 2.7 times the unresolved core estimated at 7.5 GHz in JVLA data by \citet{baldi19}, suggesting an extended steep spectrum structure.

\begin{figure}
   \centering
   \includegraphics[width=7cm]{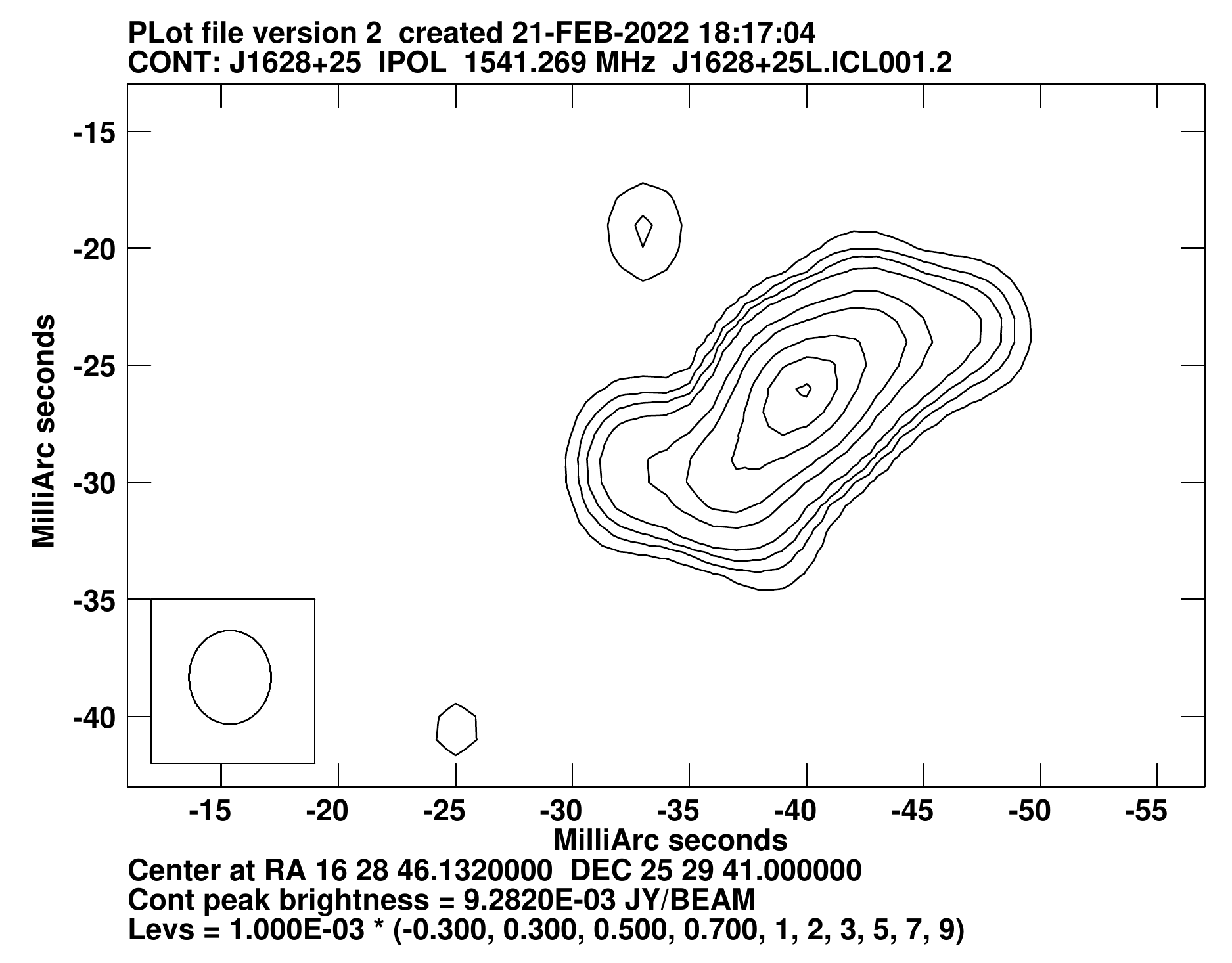}
      \includegraphics[width=7cm]{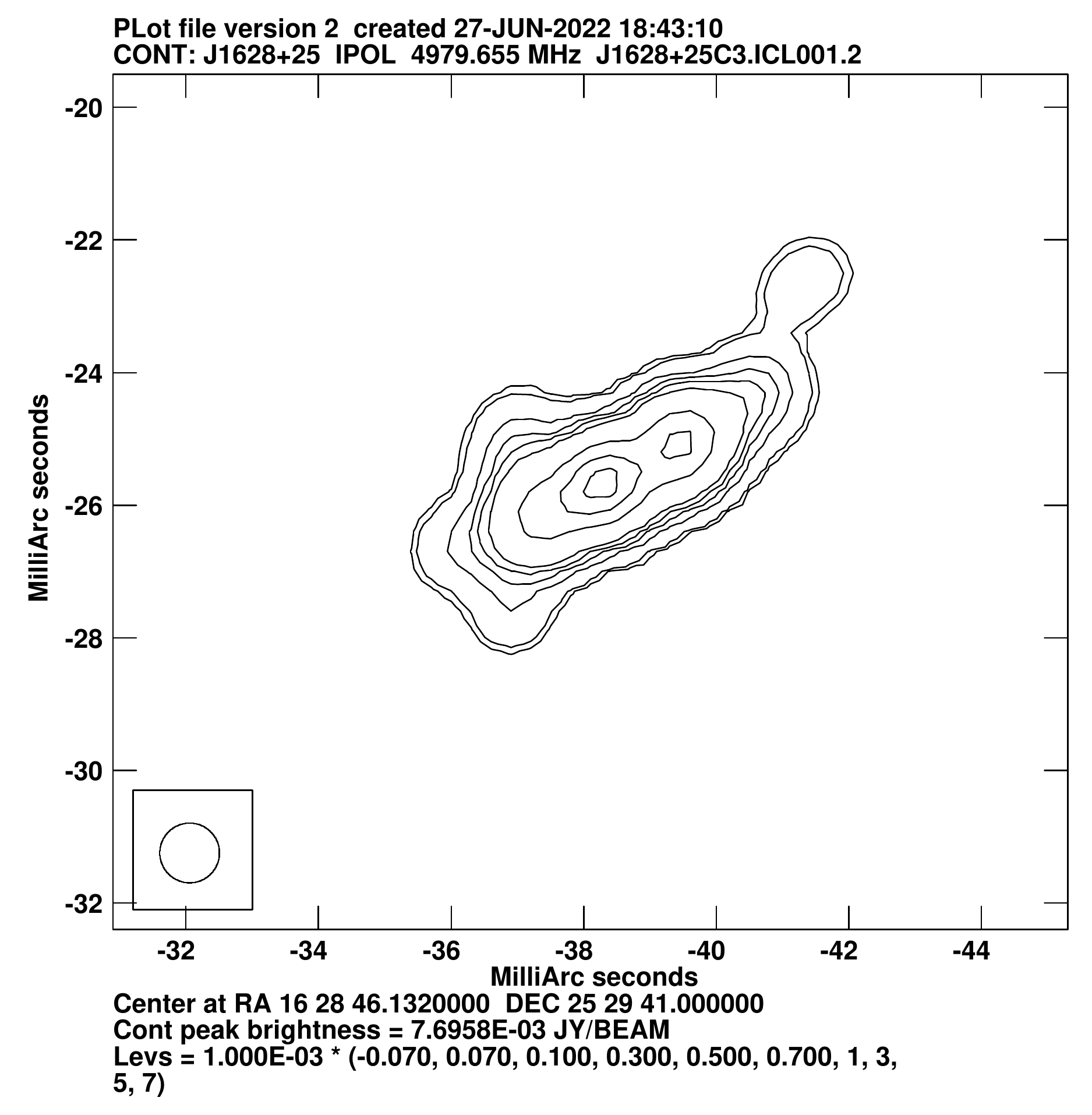}                                        
      \caption{Contour image of J1628+25 obtained with  VLBA at 1.5 GHz (top) and 1.5 GHz (bottom) 
              }
         \label{Fig12}
          \end{figure}

 {\bf J1628+25} - This source is unresolved in JVLA high-resolution images \citep{baldi19} and in the EVN image. At higher resolution (VLBA images) it is resolved in a
 symmetric structure. The inner 10 pc are in PA 115$^\circ$ (1.5 and 5 GHz image), but
 the more external regions visible only at 1.5 GHz rotate to a  PA $\sim$ 90$^\circ$ (Fig.~\ref{Fig12}). 

\begin{figure}
   \centering
   \includegraphics[width=7cm]{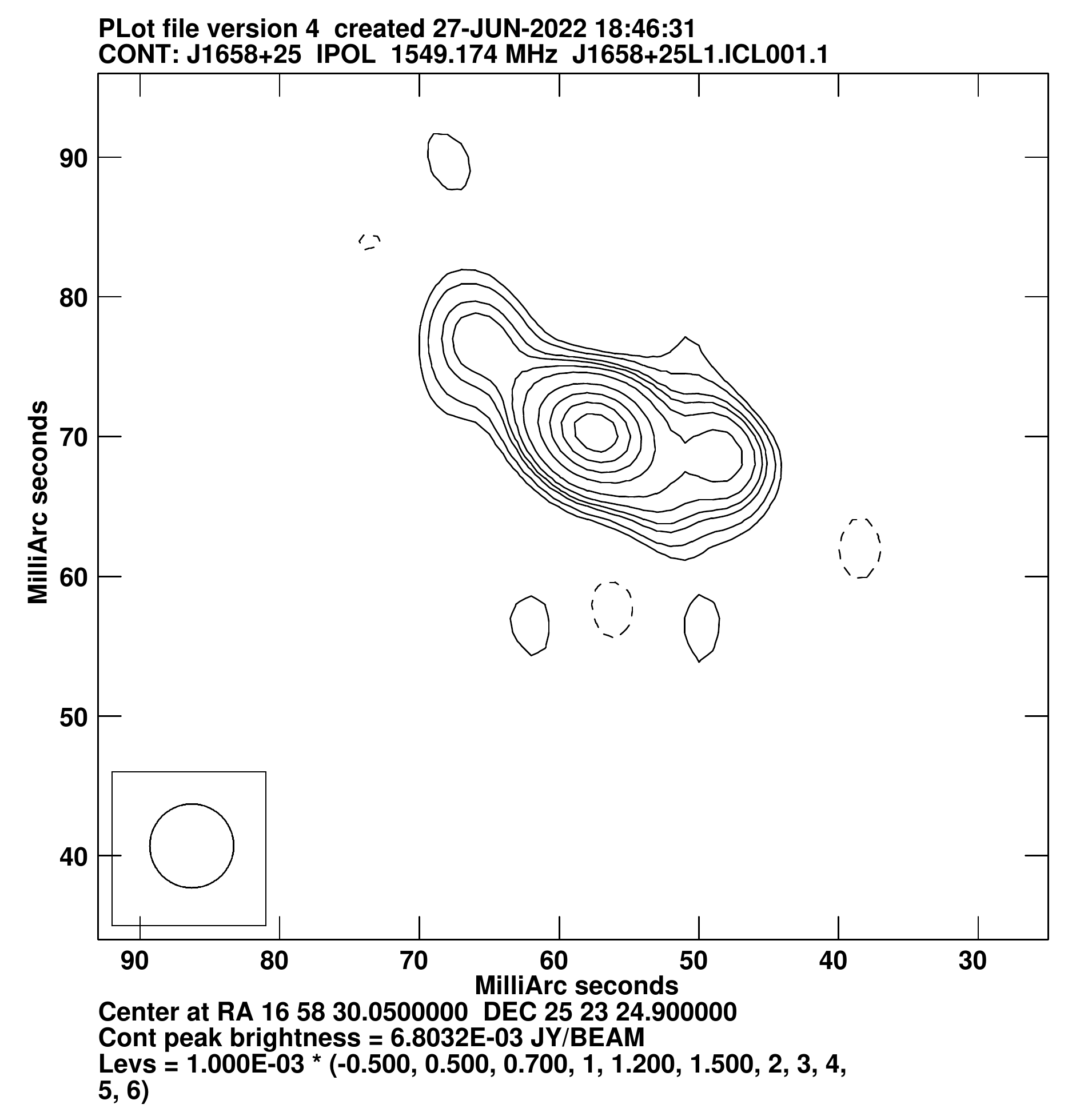}
   \caption{Contour image of J1658+25 obtained with  VLBA at 1.5 GHz.}
         \label{Fig13}
               \end{figure}

 {\bf J1658+25} - Two-sided jets are present in the VLBA 1.5 GHz image (Fig.~\ref{Fig13})
 slightly bent with respect to the inner structure. In the 5 GHz VLBA image (not shown here)
 only a slightly extended core emission is visible. The correlated VLBA flux is in agreement with the JVLA flux density by \citet{baldi19}.
 
\begin{figure}
   \centering
   \includegraphics[width=7cm]{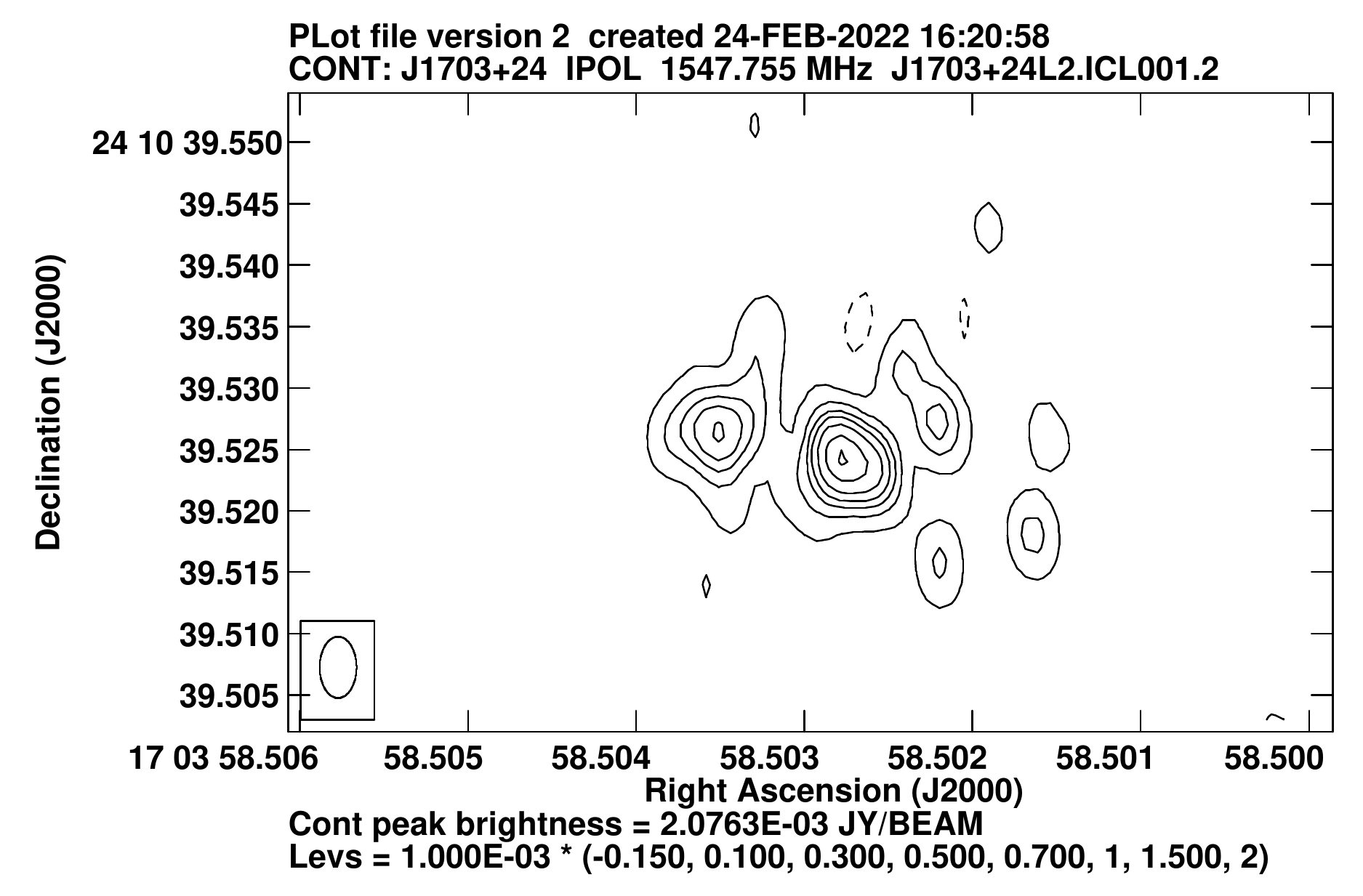}
   \includegraphics[width=7cm]{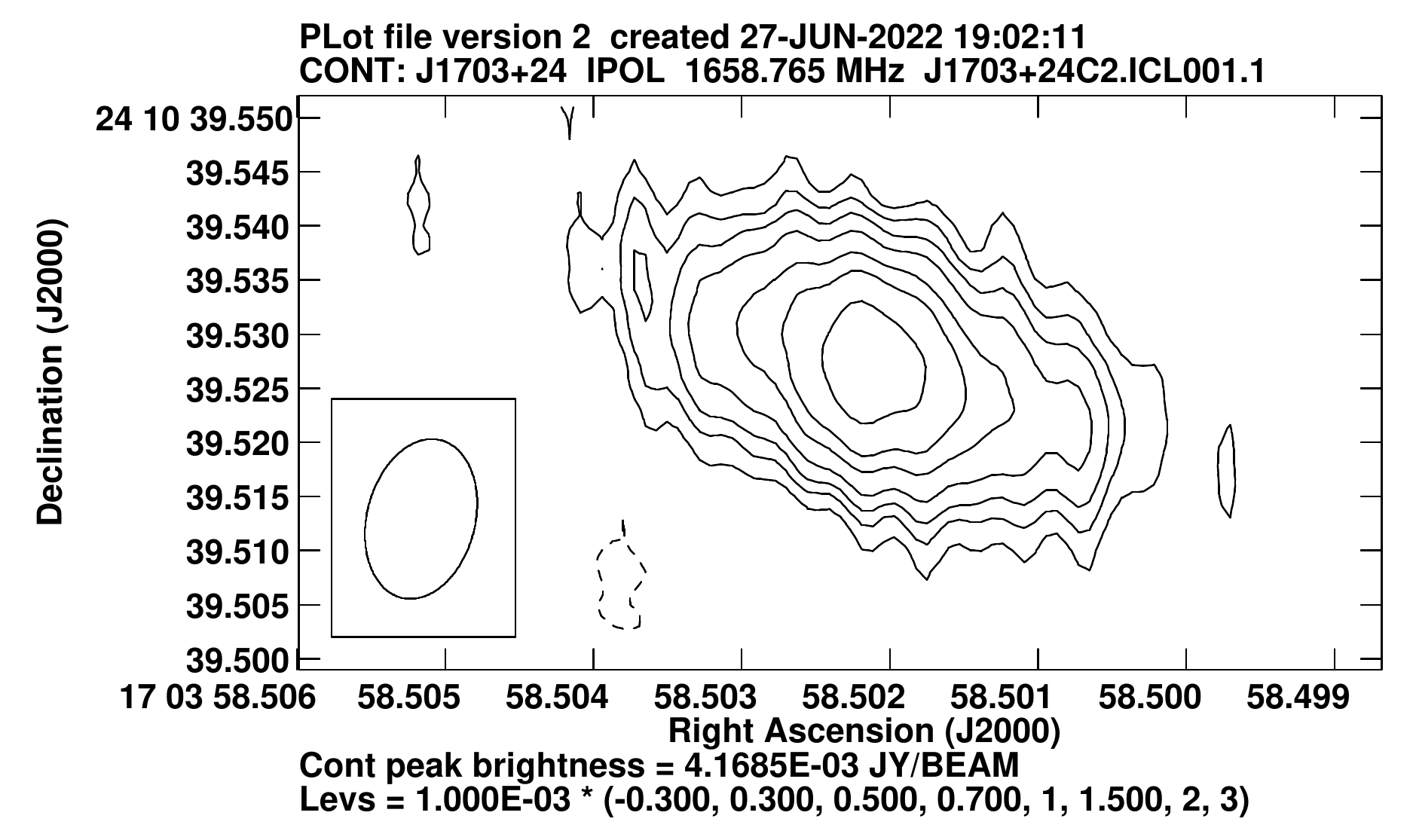}
      \caption{Contour image of J1703+24 at 1.5 GHz with  VLBA (top) and  EVN at 1.7 GHz (bottom).}
         \label{Fig14}
               \end{figure}

 {\bf J1703+24} - This is a complex source. In the  VLBA images we see a complex two-sided
 structure oriented in E-W (Fig.~\ref{Fig14}, top panel). In the EVN image a symmetric two-sided
 structure is confirmed, but at a different PA: 70$^\circ$ (Fig.~\ref{Fig14}, bottom panel).
 We note that the JVLA PA is $\sim$ 110$^\circ$
 \citep{baldi19}. The source is detected also in LOFAR observations \citep{capetti20} and it is slightly extended (5-6 kpc), in agreement with the JVLA image.

 To confirm this complex structure we self-calibrated in phase only our data at 1.5 and 1.7 GHz. Self-calibrated images have a better noise level; the source structure is confirmed. We also compared images at different frequencies at different angular resolution using the same uv-range (cutting long or short baselines). The present results are in agreement with all the tests we performed.


 {\bf J2336+00} - We did not detect this source with VLBA at 1.5 and 5 GHz.
 This could be due to a bad choice of the phase calibrator (J2340-0053),
 too faint and resolved. However, this source has been slightly resolved by eMERLIN \citep{baldi21c}, which shows a core of 0.64 mJy beam$^{-1}$ with a weaker second component. Moreover, the target source is resolved in JVLA observations \citep{baldi15} with a faint nuclear emission (1.96 mJy, at 7.5 GHz) with
 a steep spectrum: 1.00 between 1.4 and 7.5 GHz suggesting a faint resolved core.

\begin{figure}
   \centering
   \includegraphics[width=7cm]{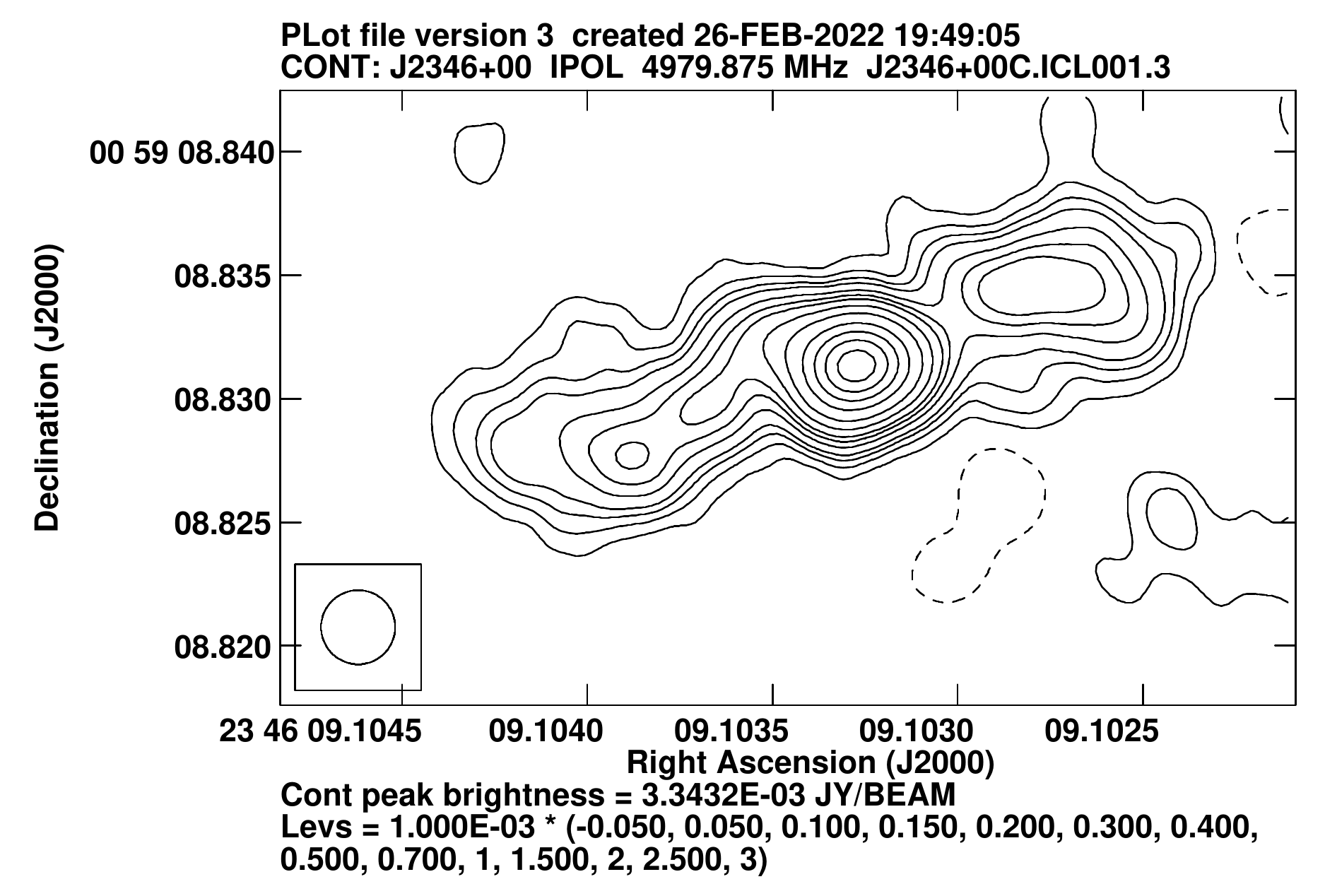}
   \caption{Contour image of J2346+00 at 5 GHz (VLBA).}
         \label{Fig15}
 \end{figure}

 {\bf J2346+00} - We observed this source with the VLBA at 1.5 and 5 GHz. The images show a two-sided symmetric emission $\sim$ 50 pc in size. The inner
 nuclear emission ($\sim$ 10pc) is at PA $\sim$ 90$^\circ$; the two-sided extended
 jet emission is at PA 105$^\circ$, suggesting a Z-shaped structure (Fig.~\ref{Fig15}).
 It is unresolved in JVLA images by \citet{baldi15}.

 \subsection {Other sources}

 VLBI observations of FR0 sources not included here have been mostly presented in \citet{cheng18} and \citet{cheng21}. We already quoted their results for the sources in common with our
 sample; here we briefly report their results for sources not present in our
 sample to increase our knowledge of FR0 parsec-scale jets.

 {\bf J0906+41} - This is a core-dominated source with a two-sided symmetric structure.
 The ratio of the NVSS flux density at 1.4 GHz to the total correlated
 flux density at 8.4 GHz is $\sim$ 1.7--1.8 suggesting an inverted radio
 spectrum.
 The jet-to-counter-jet ratio is low, 3.9, and no relativistic boosting is
 necessary. 
 
 {\bf J0909+19} - This is a two-sided source; the southern jet shows a large bend, from
 PA 180$^\circ$ to PA 110$^\circ$ (see \citealt{cheng18,cheng21}).
 The 8.4 GHz correlated flux (104.5 mJy in \citealt{cheng18} and 156 mJy in \citealt{cheng21}) is
 significantly higher than the NVSS flux density (69.1 mJy),
 suggesting an inverted radio spectrum and/or a source variability.

 {\bf J0910+18} - This is a point-like source in the  VLBA 8 GHz image \citep{cheng18}; it shows a diffuse
 low-brightness halo structure in a LOFAR image \citep{capetti20}.
 No connection is evident between the central structure
 and the halo low-brightness structure, suggesting that the diffuse low-brightness
 emission is the remnant of a previous activity. This structure shows a
 relatively steep radio spectrum ($\alpha$ = 0.96):  480.7 mJy at 150 MHz \citep{capetti20}
and 50 mJy at 1.4 GHz (NVSS).
 
 {\bf J0933+10} - This is a one-sided source. The ratio of the correlated VLBI flux
 to NVSS is only $\sim$ 0.6, suggesting a missing structure in VLBI images.
 The jet-to-counter-jet ratio is $>$ 8.

 {\bf J1037+43} - This is a one-sided core dominated structure \citep{cheng21}.
 The correlated flux is significantly lower than the NVSS source flux density
 (the ratio is $\sim$ 0.2), suggesting
 the presence of a missing structure in VLBI images and/or source variability. The jet-to-counter-jet ratio is $>$ 20.8; assuming
 that it is due to Doppler relativistic beaming, it requires a velocity larger
 than 0.54 c and an angle smaller than 57$^\circ$.

 {\bf J1205+20} - In images by \citet{cheng18} and \citet{cheng21}, this source shows a
 one-sided jet and a diffuse lobe-like structure on the other side. The
 correlated flux is 41.2 mJy at 8.4 MHz and the NVSS flux density is 89.9 mJy.
 The jet-to-counter-jet ratio near the
 core is $\sim$ 33 (v $>$ 0.6c and $\theta$ $<$ 50$^\circ$). The spectral index of
 the diffuse NE structure is 0.75 or steeper. It could be the remnant of a
 previous activity and/or a peculiar structure related to the counter-jet
interaction with the ISM.

 {\bf J1246+11} - This is a one-sided jet structure in \citet{cheng18}. The source
 flux density in the NVSS is 61.2 mJy, 38 mJy in \citet{cheng18} at 8 GHz and
 only 7--8 mJy in \citet{cheng21} at 5 and 8 GHz. More observations are necessary to understand this source.

 {\bf J1604+17} - Only a core-like structure is visible in the image reported
 by  \citet{cheng18}, even if in the table it is reported as a jet-like structure.
 The correlated
 flux is in agreement with the NVSS flux density measure for this source. A
 tail-like structure interacting with a nearby galaxy is present in the LOFAR
 image in \citet{capetti20}.

 {\bf J1606+18} - A one-sided source is reported in  \citet{cheng18}. The correlated flux
 density is 360 mJy; the NVSS flux density is 396 mJy.
 The   jet-to-counter-jet ratio is $>$ 8.
 
 {\bf NGC5322} - This source at z= 0.00594 was studied by \citet{feretti84}
 and more recently by \citet{dullo18}. The
 source shows a two-sided symmetric extended jet structure.

 {\bf NGC1052} - This source at z = 0.00496 is a well-known active radio galaxy
 because of its low redshift. \citet{backzo22} confirms the presence of a
 remarkably straight low-velocity double-sided jet.
 
 \section{Discussion}

 The low radio power and the lack of a kiloparsec-scale radio emission in FR0 sources could suggest an origin of the radio emission different from the classic non-thermal
emission from a SMBH and jets, for example star formation or a disc--corona wind 
emission, similarly to what has been suggested for radio quiet AGN (e.g., \citealt{panessa19}). 
The widespread detection of compact parsec-scale jets in FR0s indicates that  the origin of the radio emission is synchrotron emission from outflows produced by a SMBH at the center of the parent galaxy.

\subsection{The jet-to-counter-jet ratio}
  
  The jet-to-counter-jet ratio, $R_{\rm JC} =F_{\rm jet}/F_{\rm cjet}$, can be ascribed to the relativistic boosting of two otherwise symmetric jets  (see, e.g., \citealt{giovannini04}). Therefore, this parameter can be linked to the jet bulk speed, and it is very useful to study the jet velocity distribution.
To this end, \citet{giovannini05} selected a sample of radio galaxies from the B2 Catalogue of Radio Sources and the Third Cambridge Revised Catalog (3CR), forming the Bologna Complete Sample (BCS), with no selection constraint on the nuclear properties. 
The   jet-to-counter-jet ratio distribution of the BCS sources was discussed by \citet{liuzzo09}, who found that $\sim$ 30\%
 of the sources show evidence of a two-sided structure. This result is in agreement with a jet velocity
of $\beta \sim$ 0.9, and a random orientation of radio galaxies with respect to our line of sight. 
 
 \begin{figure}
   \centering
   \includegraphics[width=9cm]{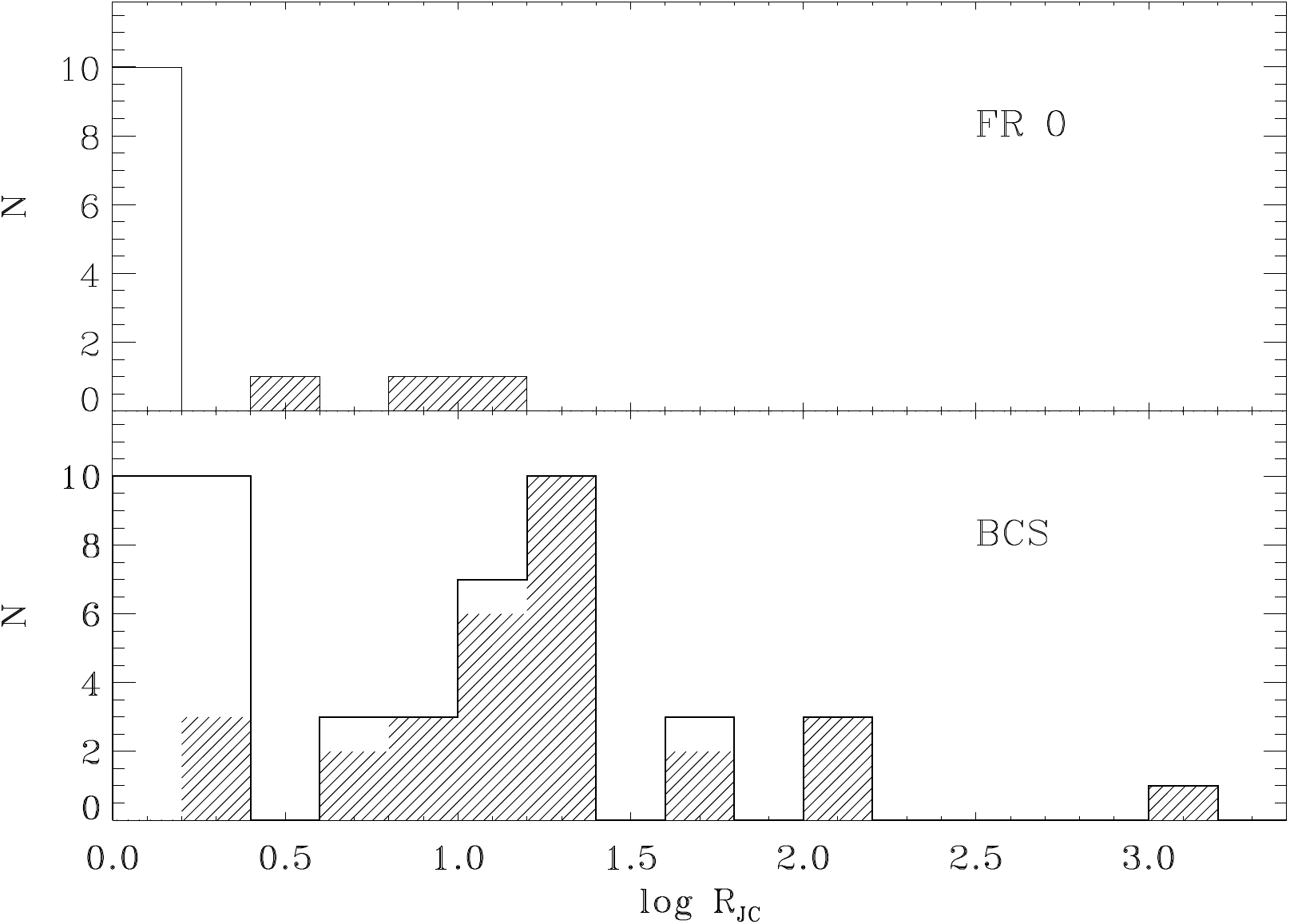}
   \caption{Distributions of the logarithm of the ratio R$_{\rm JC}$ between the flux densities of the jet and counter-jet for the FR0s (top panel) and the source in the BCS sample (bottom panel). The dashed portions of the histograms correspond to lower limits on R$_{\rm JC}$.
         }
\label{hist}
 \end{figure}
 
 In new sources presented here and in the literature data (see Sects. 3.1 and 3.2), we
 find that 56\% of FR0s show a two-sided structure
 (52\% if we exclude NGC5322 and NGC1052). The FR0 sample with VLBI observations is not a complete sample, but it is
 representative of the parsec-scale structures of this class of sources if
 we take into consideration selection effects. Only sources with an unresolved nuclear emission in JVLA images larger than 20 mJy at 1.4 GHz (see Sect. 2) were selected for EVN observations. Moreover, sources observed by \citet{cheng18} have a higher nuclear flux density with respect to the general FR0 population. 
These selection biases in our sample favor sources whose nuclear emission is boosted by Doppler beaming, that is, those seen at a smaller angle with respect to the line of sight. Therefore, we would expect as selection effect a lower percentage of two-sided structures with respect to a sample of randomly distributed sources as the BCS. Conversely, the percentage of two-sided sources (56\%) found here is significantly higher than the 30\% found in the sample by \citet{liuzzo09}. 

We conclude that the high percentage of FR0 with a two-sided structure is  strong evidence that parsec-scale jets in FR0 sources are mildly relativistic or not 
 relativistic (v $\sim$ 0.5c or lower).
  The lower bulk speed of the jets in FR0s with respect to the BCS sources is also supported   by the distribution of the jet-to-counter-jet ratio for the two samples presented in Fig.~\ref{hist}. The fraction of FR0s with $R_{\rm JC}> 2$ is 23\%, while this fraction is between 60\% and 66\% (depending on the actual values of the lower limits) in the FRIs. By considering the upper limits, the mean values of the two distributions (Kaplan Meier estimator) are 2.7$\pm$1.0 and 1.2$\pm$0.2 for BCG galaxies and FR~0s, respectively. The probability that the two populations are drawn from different distributions of $R_{\rm JC}$ is $>$98\%.\footnote{The tests were performed using a censored statistical analysis with the task twosampt within the ASURV package \citep{lavalley92} implemented within the PyRAF software \citep{pyraf12}.} Furthermore, because the FR0s sample is biased toward sources seen at a smaller angle with respect to the line of sight, this difference is likely to be underestimated. Unfortunately, the incompleteness of the FR~0s sample prevents us from deriving a robust estimate of their jet speed.
 
 \subsection{Jet morphology}

 The jet morphology in many sources presented in the previous section is
 complex. Many FR0 jets are transversely resolved, with blobs and
  substructures suggesting a strong interaction with the surrounding ISM. 
 One of the best case is J1040+09 (Fig. 7) where at 5 GHz we see a northern jet
 with an S-shaped structure, and at 1.5 GHz we have a well-resolved two-sided structure. 
 The southern one is more extended and is lobe-like extended
 to the east. 
 
  These features are instead not seen in the inner region of FRIs. This difference can be interpreted as being due to  different jet speeds in the two classes: the slower and less energetic jets of FR0s are more affected by the presence of a nonhomogeneous external medium; the relativistic jets of FRIs are instead able to penetrate it without being deflected. Only at a distance from the jet basis of several kiloparsec, when the jet velocity
 decreases, do FRIs show distorted and complex morphologies.

Three FR0s, namely J1621+25, J1628+25, and J2346+00, show a Z-shaped structure. The jets are symmetric very near to the core, 
 and move to a different PA after a few milliarcseconds at about the same distance where they form symmetric extended structures at a different PA. A similar structure has been rarely found in FRI sources and on a larger
 scale. The best cases are 3C 293  \citep{vanbreugel84} and 4C 26.42 \citep{liuzzo09b}. These structures have been interpreted as the
 interaction between a low-velocity jet and a surrounding rotating gas disk
 structure (3C293) or as a low-velocity jet interacting with
 rotating accretion gas at the center of the cooling
 cluster A1795 \citep{liuzzo09b} for 4C 26.42.
 This suggests that   in FR0 sources the Z-shaped structure could also be due to a gas rotating disk-like structure coupled with a low-velocity jet.

 We conclude that   the FR0 morphology found in available VLBI images also show  strong
 evidence that parsec-scale jets in FR0 sources are not relativistic or at most are mildly
 relativistic. 

 \subsection{Missing correlated flux}

 We compared the correlated VLBI flux with the nuclear emission detected in
 JVLA images
 at 7.5 GHz, when available, and with other VLA images in other
 cases. In 40\%
 of FR0 sources we find an agreement between the two results suggesting that
 VLBI images
 show all or most of the source nuclear structure. In a lower number (20\%) we
 found a
 higher flux density in VLBI images, possibly due to source variability and/or an inverted or peculiar radio spectrum.
 In other sources (40\%)  a significant fraction of flux density is missing in
 VLBI images with
 respect to the arcsec-scale images. This loss can be due to
 variability, but mainly to the lack of
 sensitivity in VLBI images to diffuse emission; the VLBI images
 resolve and
 do not see low-brightness extended structures because of super-resolution and spectral index effects. This is also evident when  comparing high- and low-frequency VLBI images (see, e.g.,  the EVN images at 1.7 GHz of J1230+47 with the
 8.4 GHz of the same source by \citealt{cheng18}). A deep study of the sources where a significant fraction of their flux density is missing in VLBI images will allow us to understand why jets stop and die. To understand jet properties we need to understand how the jets evolve. The FR0 jets could be able to create a diffuse low-brightness lobe-like emission or an extended low-brightness jetted structure, both not visible in VLBI images. The interaction between the host galaxy and the nonthermal jet emission would be different in the two cases. Moreover, in the one-sided sources the missing flux could be due to the invisible counter-jet emission.
 Intermediate-scale observations
 (e.g., with EVN and eMERLIN) and the combination of long and short baselines (as demonstrated in \citealt{baldi21c} with eMERLIN and JVLA combined data sets) represent a powerful tool to study the missing flux, for example    in a lobe-like structure created by the jet,   in a longer jet becoming increasingly fainter 
 moving far from the core, or   a contribution of the counter-jet side in asymmetric sources. 
 
\subsection{Origin of the radio emission}

Theoretical models indicate that an accreting SMBH is the origin of symmetric relativistic jets (see, e.g., \citealt{blandford79,gomez95} and references therein) in contrast with the low-velocity jets found here. However, observational evidence of low-velocity jets cannot exclude the presence of an inner relativistic spine surrounded by a low-velocity sheath. The main difference between FRI and FR0 jets could be the relative importance of the relativistic spine in the jet structure.

The present data are in agreement with the \citeauthor{garofalo19} model where
low-power relativistic jets are expected in sources with a low prograde spinning black hole surrounded by advection dominated accretion flows. In these sources the relativistic inner spine of the jet launched by the black hole ergosphere associated with the Blandford-Znajek process \citep{blandford77} is less prominent than that produced by the Blandford-Payne portion of the accretion disk \citep{blandford82}. 

In this framework we expect a jet
with a subtle relativistic spine unable to extend, in most cases, to the kiloparsec scale due to instabilities. This jet is
surrounded by a low-velocity sheath, which is the dominant
structure visible in our images. This jet structure accounts for the nature of the
very few compact FR0s,  also in VLBI images (e.g., J0943+36) where the structure, the high flux density, and the high variability suggest that we are seeing a relativistic jet
well aligned with the line of sight. 

FR0 jets, because of the large diffuse sheath, are influenced by the ISM, but also strongly interact with it  and 
 modify its properties and physical conditions. Mapping HI on  parsec scales 
 \citep{murthy22} show that  a low-power jet is also able to remove cold gas 
 from the center of a galaxy, and therefore to influence the accretion into the
 SMBH and the properties of the host galaxy. The large number of FR0 galaxies
 implies that the feedback between the energy produced by the AGN and the galaxy evolution is also present  in sources where the total radio power is low because
 the low-velocity jet is not able to create and sustain extended radio lobes.

 \section{Conclusions}

FR0 galaxies are the dominant local population in the radio sky. They have multiwavelength nuclear properties similar to FRI radio galaxies, but show a clear deficit of extended radio structures and are largely core dominated with respect to FRI sources. Most FR0s are unresolved in radio images at arcsecond resolution. For this reason VLBI observations are necessary to investigate the nature and properties of their radio emission.

From the new observations presented here and literature data we conclude the following:

\begin{enumerate}

\item All FR0 sources observed with VLBI, with the exception of J2236+00, have
 been detected. This result, together with the observed jet structures visible in most
 VLBI images, show that the FR0 radio emission is related to the nuclear activity of an accreting SMBH from where two symmetric synchrotron jets are launched.
 
\item The complex jet
structures and the large number of two-sided structures are 
strong evidence that the FR0 jets present in VLBI images are not relativistic, contrary to what is observed in FRI sources.

\item The properties of a few FR0s also suggest
the presence of a relativistic faint inner spine in these sources, as requested by the proposed identification as misoriented BL Lacs. 
These results are in agreement with the suggestion that FR0 jets are associated with a low (and prograde) spin SMBH. In these sources jets are structured; a relativistic faint inner spine launched by
the black hole ergosphere (BZ jet) is surrounded by a dominant nonrelativistic shear launched by the accretion disk (BP jet). 
Because of the strong interaction between the shear and the ISM, the
jet cannot survive to reach the kiloparsec scale.
 
\item In several FR0 galaxies the correlated parsec-scale flux density is a small fraction of the nuclear flux density measured at subarcsec resolution, indicating that a significant amount of radio emission is produced at intermediate scales. EVN and MERLIN observations are necessary to understand and probe this component, most likely associated with low surface brightness jets or diffuse lobes. 

\item Despite  the low power and small size of radio emission in FR0s, our results represent clear evidence for the presence of collimated jets. This implies that AGN feedback is also at work  in these sources. Recent observations have shown that a low-power jet is able to remove cold gas from the center of a galaxy, and therefore to influence the accretion onto the SMBH.
 
\end{enumerate}




\begin{acknowledgements}
We would like to thank the Referee for useful comments.
RL acknowledges financial support from the State Agency for Research of the Spanish MCIU through the “Center of Excellence Severo Ochoa” award for the Instituto de Astrofísica de Andalucía (SEV-2017-0709), from the Spanish Ministerio de Economía y Competitividad, and Ministerio de Ciencia e Innovaci\'on (grants AYA2016-80889-P, PID2019-108995GB-C21), the Consejería de Economía, Conocimiento, Empresas y Universidad of the Junta de Andalucía (grant P18-FR-1769), the Consejo Superior de Investigaciones Científicas (grant 2019AEP112).

The National Radio Astronomy Observatory is a facility of the National Science Foundation operated under cooperative agreement by Associated Universities, Inc.

The European VLBI Network is a joint facility of independent European, African, Asian, and North American radio astronomy institutes. 
\end{acknowledgements}

%
   \bibliographystyle{aa} 
%

\bibliography{main}  






   
  



\end{document}